\documentclass[preprint2]{emulateapj}
\usepackage{natbib,amsmath, amsfonts, amssymb}

\begin{document}

\def\kms{${\rm km~s^{-1}}$}
\def\mkms{{\rm km~s^{-1}}}
\def\jgalx{J204956.61$-$001201.7}
\def\mhkpc{h_{72}^{-1} \, \rm kpc}
\def\hkpc{$h_{72}^{-1} \, \rm kpc$}
\def\qsonm{J2049$-$0012}
\def\BOL{ Bol}

\title{Shining Light on Merging Galaxies I:\\ The Ongoing Merger 
of a Quasar with a `Green Valley' Galaxy
}

\author{
Robert L. da Silva\altaffilmark{1,2},
J. Xavier Prochaska\altaffilmark{1}, 
David Rosario\altaffilmark{1}, 
Jason Tumlinson\altaffilmark{3},
Todd M. Tripp\altaffilmark{4}
}
\altaffiltext{1}{Department of Astronomy and Astrophysics, UCO/Lick Observatory, University of California, 1156 High Street, Santa Cruz, CA 95064}
\altaffiltext{2}{NSF Graduate Research Fellow}
\altaffiltext{3}{Space Telescope Science Institute, 3200 San Martin Dr., Baltimore, MD 21218}
\altaffiltext{4}{Department of Astronomy, University of Massachusetts, 710 North Pleasant 
Street, Amherst, MA 01003-9305}

\begin{abstract} 
Serendipitous observations of a pair $z = 0.37$ interacting galaxies
(one hosting a quasar) show a massive gaseous bridge of
 material connecting the two objects. This bridge is photoionized by
 the quasar (QSO) revealing gas along the entire projected 38 \hkpc\
 sightline connecting the two galaxies. 
The emission lines that result give an unprecedented opportunity to
study the merger process at this redshift.  
We determine the kinematics, 
ionization parameter ($\log U \approx -2.5 \pm 0.03$), column density
($N_{H,\perp}\approx 10^{21} \mbox{ cm}^{-2}$), 
metallicity ([M/H]$\approx-0.20\pm0.15$), and mass 
($\approx 10^8 M_\odot$) 
of the gaseous bridge. We
simultaneously constrain properties of the QSO-host ($M_{DM}>8.8\times
10^{11}$) and its companion galaxy ($M_{DM}>2.1\times 10^{11}$;
$M_\star\sim2\times10^{10} M_\odot$; stellar burst age=$300-800$~Myr;
SFR$\sim6\ M_\odot$ yr$^{-1}$; and metallicity $12+\log$(O/H)=$8.64\pm0.2$). 
The general properties of this system match the standard paradigm of
a galaxy-galaxy merger caught between first and second passage
 while one of the galaxies hosts an active quasar. 
The companion galaxy lies
in the so-called `green valley', with a stellar population 
consistent with a recent starburst triggered 
during the first passage of the merger and has \emph{no}
detectable AGN activity. 
In addition to providing case-studies of quasars associated with galaxy mergers,
quasar/galaxy pairs with QSO-photoionized tidal bridges such as this one
offer unique insights into the galaxy properties while also
distinguishing an important and inadequately understood phase of galaxy evolution.
\end{abstract}

\section{Introduction}
The paradigm of hierarchical structure formation is based on the
frequent mergers of galaxies and their dark matter halos, as
predicted and quantified
 by cosmological simulations which construct merger trees to track the merger history of halos
 \citep[e.g.][]{millennium,mergertrees}.  Observations of galaxies `caught in the act' of merging \citep[e.g.][]{arp, lotz08} support
 this theory. 
Galaxy mergers are believed to be an important phase of galaxy
evolution and are thought to be crucial to understanding 
super-massive black hole (SMBH) growth, galaxy morphologies, and the truncation of star-formation
within galaxies.

While details of the merger process 
 depend on the orbit, mass ratio and morphologies of the galaxies involved, the basic
sequence of events for a major (galaxy mass ratio $\gtrapprox$ \ 1:4) is relatively 
well agreed upon \citep{mihos1994b, mihos96, hop08, cox2006,
  dimatteo2007,galmer}.
Galaxies undergoing a merger typically have orbital angular momentum that prevents
direct collisions and leads to a series of close
passages which have progressively smaller apogees due to dissipation
in the merger. 
During their first close passage, the galaxies exert
strong tidal forces on each other. For galaxies with a small bulge component, 
these tidal forces lead to
bar formation. The stellar bar lags behind the gaseous bar inducing
a torque on the gas which leads to the funneling of gas towards the
center of the galaxy.
The inflow also leads to higher central gas surface densities which is
predicted to spur
 enhanced star formation \citep[e.g.][]{mihos96}. 

 Subsequent close passages lead to smaller inflows as a large fraction
 of the available gas 
  was used up during a first-passage `starburst'. The second major
 phase of inflow, therefore, occurs later due to a different mechanism
 as the galaxies are engaged in final
coalescence. At that time, the remaining gas is brought rapidly to the center
 by dissipation into a dense compact central 
structure that rapidly forms stars.
The final remnant galaxy in merger simulations 
is commonly an elliptical, dispersion-dominated system
that has ceased star-formation because its gas reservoir is used up in
the starbursts and lost via stellar or AGN feedback \citep[e.g.][]{springel05}. 
However, the details of the effectiveness of `quenching'
star-formation and
the actual process that is responsible varies between different studies and depends on the choice
of numerical prescription.  

For galaxies with large bulge components, the initial bar formation is
suppressed and star formation is far less elevated in the early stages
of the merger. Consequently,  there is a large reservoir of gas
remaining until the stages of final coalescence and the final
starburst is more extreme. Because the galaxies coalesce over a short
timescale, the final starburst can be
more intense than that of a bulgeless galaxy
during first passage.

Some studies \citep{hop08,springel05} argue that these merger events lead to elevated
levels of AGN activity and ultimately a quasar (QSO) phase which is
short \citep[$10^6-10^{8.5}$ yrs;][]{mart01, mart03} and extremely
bright ($\sim10^{46}\,\mbox{erg/s}$). This phase of the supermassive black hole
(SMBH) evolution is thought to dominate the mass accretion history of
the black hole \citep{soltan} and can profoundly affect its host.
During final coalescence, the large gas densities near the center of the galaxy (where the SMBH
is located) leads to a large amount of efficient AGN fueling that manifests as a QSO phase.
This scenario is supported by observations of quasar hosts showing disturbed
morphologies \citep[e.g.][]{ben08, green2010} consistent with being merger remnants.
Observations disagree on whether mergers lead to elevated 
levels of AGN activity in the early stages of the merger (i.e. between first and
second passage)
 with some \citep[e.g.,][]{woods}  suggesting an increase and others \citep[e.g.,][]{ellison}
suggesting no increase in activity at these early stages.

In addition to their obvious role in increasing galaxy mass, 
mergers provide a tantalizing solution to several observed cosmological trends.
Firstly, they offer a convenient explanation
for the various SMBH-host relations. Na\"{i}vely, the SMBH (which only has a gravitational
sphere of influence $\approx 1/1000$ the size of the host) should have no correlation
with the properties of its host. However, SMBHs have been observed to be correlated with
their host's stellar mass \citep{mag98}, velocity dispersion \citep{geb2000,ferrarese2000}, dark matter halo mass \citep{beyondbulge},
and morphology \citep{graham2001}. If black holes are growing in galaxy mergers at the same
time that many of these other galactic properties are changing, then
this may explain the observed correlations with the host. In fact, it
has been proposed that this simultaneous evolution can lead to these various
correlations even without invoking any AGN feedback \citep{sutter2010}.

Mergers may also be a key ingredient in quenching star-formation
within galaxies.
The basic picture is that galaxies can be broadly classified into one
of two groups: (1) actively star forming spiral ``blue cloud" galaxies
or (2) passively evolving elliptical ``dead"  red sequence galaxies.
 It appears that the red-sequence galaxies somehow evolve from
actively star-forming galaxies \citep[see ][]{faber07}. 
These
hypotheses raise the 
question of how 
actively star-forming galaxies are ``terminated" to produce red-sequence galaxies. While there
have been many theories proposed \citep[e.g., halo quenching;][]{dekel},  one of the
more popular solutions is through the AGN activity, starbursts, and
feedback induced by galaxy mergers \citep[e.g.][]{hop08}. In such a
paradigm, the sequence of events (as described above) of the merger
lead to a burst of star formation and AGN activity whose
feedback can unbind a large fraction of the remaining gas. This quenches the
galaxy while the merger also leaves the remnant with an elliptical
morphology, seemingly consistent with observations \citep{toomre77}.

This theory is not without its flaws. Observational evidence suggests that 
galaxies transitioning from the ``blue cloud" to red sequence (i.e. ``green valley"
galaxies) do not appear to be dominated by merger remnants \citep{reichard} nor the result
of rapid quenching \citep{martin2007}.  Nevertheless, it has
established a theoretical framework that, in principle, can be tested
with empirical observation.

 The details of SMBH fueling and feedback as well as star formation and supernova feedback
 are crucial to understanding the actual role that mergers play in galaxy evolution.
However, merger simulations cannot realistically implement these processes due to
  the relevant physical scales spanning a dynamic range of sub-pc to
  many kpc scales. 
    Additionally, simulations are regularly unable to accurately
  couple radiative transfer to the hydrodynamics due to high computational expense.
 Simulators are thus forced
  to implement tuned sub-grid models which sometimes are not physically well-motivated 
  (e.g., using Bondi-Hoyle accretion; see \cite{booth2009}). 
  Thus there remain many open questions about
  the processes at work during a galaxy merger that simulations will be unable to enlighten 
  for the foreseeable future. 
As such,
observations are required to improve our understanding and
guide future theoretical effort.
However, 
observational studies of these processes are challenging for a number of reasons. 
In contrast to studies of 
nearby, merging galaxies which permit sensitive searches for
disturbed morphologies and faint
   tidal structures \citep[e.g.,][]{arp}, such analysis at higher redshift
   becomes prohibitive owing to the great distance.
  Low surface brightness tidal 
   features are too faint to observe and one requires much higher
   angular resolution
   to detect small morphological features at higher redshift.
Additionally, for all but the nearest galaxies 
\citep[e.g.,][]{hib01}, observational studies of mergers 
are limited to studying the
stellar distribution because the emission from the bulk of the gas is
invisible at cosmological distances.  However, it has been shown that
gas and stars may have completely different velocity structures and distributions,
especially in systems where a galaxy is
undergoing a phase of enhanced AGN activity \citep{gre09}. Since the
dynamics of the gas and not the stars ultimately fuel AGN activity and star formation,
meaningful studies of these processes require the ability to study gas
in mergers. 
  Lastly, while the various feedback prescriptions may be tuned
   to match global properties of an ensemble population of galaxies, they
  must also be tested against actual merging systems.

This gap in our understanding motivated a novel method for finding and
studying merging galaxies that is inspired by the serendipitous
discovery of a remarkable prototype. In our method, we search for
systems where the ionizing flux of the quasar ``lights up'' the gas
in a merger in order to find new observational constraints on 
merger models.

With an ionizing photon flux 
($S$) of $\approx 10^{56}$ photons/s/sr \citep{tad96},  a quasar 
can ionize gas to conditions comparable to \ion{H}{2} regions out to
distances of 60~kpc. While observations of highly ionized gas with
large (tens of kpc) spatial extent known as extended emission line
regions (EELRs) have been known for decades \citep{mat63, wamp75,
  stock87, pens90, fu09}, our study focuses on a novel application of
this standard result where we utilize quasar photoionization to study
a pair of 
merging galaxies. In detail, we focus on an EELR
produced by a quasar that has been triggered by a merger event. During
such a merger, one also expects extended tidal features of gas out to
distances of ~100 kpc that may be photoionized by the quasar and could
thus be rendered visible through recombination and forbidden
lines. Observations have confirmed that such extended material that
may be associated with a galaxy merger can be seen in emission lines
near at least one quasar \citep{villarmartin10}. 

Our prototype was serendipitously discovered while selecting targets for a
 study of gas in the extended halos of galaxies via quasar sight lines (Tumlinson et al in prep.).  
That study used the Sloan Digital Sky Survey DR5 \citep{sdssdr5} to
find foreground galaxies close to quasar sight lines based on
photometric redshifts. However the galaxy they had hoped to be in
front of J2049-0012 (a z=0.369 quasar) by a cosmological distance was
in fact found to have the same redshift. 
Further inspection revealed that the long-slit spectrum of the system
contained bright emission lines along a remarkable gas 
``bridge'' connecting the quasar\footnote{ 
  We reserve the terms `quasar' and `host galaxy' to refer to the
  galaxy that is undergoing a quasar phase while terms with the
  adjective `companion' refer to the other galaxy with which the QSO host is
  interacting. We interchange the use of the terms QSO and quasar and
  make no distinction between the two classes of objects for the purposes of
  this paper. }
to its companion galaxy, 38 \hkpc\ away in
projection, and photoionized by the hard spectrum of the quasar. This
bridge is most likely the result of a tidal interaction of the
galaxies during the first passage of their ongoing merger. 
The extended gas around these objects, normally
completely invisible at this distance, is being lit-up by radiation
from the quasar. This provides an unparalleled opportunity to study the
kinematics and structure of the gas in the merger, which otherwise 
we would not have even known was occuring. At the same time,
we have detailed measurements of the galaxy properties,
including the outer regions of the quasar's host galaxy. 
The companion galaxy is also a member of the `green valley'
and we are afforded the chance to examine 
this poorly understood and rare phase of galaxy evolution in this
merger context.  

This is the first in a pair of papers on this particular system. This
paper focuses on the details of the merging system (mass ratio,
timescales, separation, merger stage, companion galaxy properties, 
SMBH mass and luminosity) as
well as discussion of the emission line bridge and an explanation for
its ionization source. 
Paper II focuses on the quasar
including its lifetime, isotropy, impact on companion galaxy,
implications for its triggering, and the placement of the SMBH on
Magorrian-like relations at a peculiar phase of its evolution.

We assume a cosmology with $\Omega_\Lambda =0.73$, $\Omega_M =0.27$,
and $H_0 = 72\ h_{72} \rm km \, s^{-1} \, Mpc^{-1}$.

The paper is organized as follows: $\S$~\ref{sec:obs} describes the observations of this system; $\S$~\ref{sec:ovr} gives an overview of the results of the observations; $\S$~4 details our measurements of emission line fluxes and kinematics; 
$\S$~5 describes the inferred properties of the interacting pair of galaxies; $\S$~6 discusses our analysis of the bridge connecting the galaxies including our analysis of its origin; $\S$~7  presents further discussion; $\S$~8 presents a summary.


\begin{deluxetable*}{cccccccccc}
\tablewidth{0pc}
\tablecaption{Observations}
\tabletypesize{\footnotesize}
\tablehead{\colhead{Instrument} & \colhead{Slit}& \colhead{Dichroic} & \colhead{Camera}  &\colhead{Disperser\tablenotemark{c}} & \colhead{Resolution \tablenotemark{d} [km/s]} & \colhead{Filter} & \colhead{Exposure} &\colhead{Standard Star}& \colhead{Date } \\ & [$''$] & & & & & [s]& & (UT)} 
\startdata

Keck/LRIS & 1.0 & D560 & R & 600/7500 (28.15) & 160 &\nodata &900x2 & Feige 110\tablenotemark{a}&2008-10-03 \\
Keck/LRIS & 1.0 & D560 & B & 600/4000 &  240 &\nodata &  830x2 & Feige 110\tablenotemark{a}&2008-10-03 \\
Keck/LRIS & 1.0 & D560 & R & 600/7500 (29.15) & 160 &\nodata & 300x3 & BD+28\tablenotemark{b}&2009-09-17 \\
Keck/LRIS & 0.7 & D560 & R & 1200/7500 (40.15)& 65 &\nodata &600x2 & BD+28\tablenotemark{b} &2009-09-17\\
Shane/Kast & 1.0 & d55 &  R & 600/7500 (10775.6)&\nodata & \nodata &  1800 & \nodata&2009-05-26 \\
Keck/LRIS & \nodata & D560 & B & \nodata &\nodata& B & 60,200,230x2&\nodata& 2008-10-03  \\
Keck/LRIS & \nodata & D560 & R & \nodata &\nodata& R & 90,200x3&\nodata& 2008-10-03  \\
\enddata
\tablenotetext{a}{\citep{bohlin96}}
\tablenotetext{b}{\citep{bohlin01}}
\tablenotetext{c}{Values in parenthesis denote the grating angle.}
\tablenotetext{d}{Measured as the FWHM of a sky line.}
\label{table:obs}
\end{deluxetable*}

\section{Observations \& Data Reduction}
\label{sec:obs}
We obtained a set of spectroscopic and imaging observations of this quasar/galaxy
pair using the dual-channel Low Resolution Imaging Spectrometer \citep[LRIS;][]{lris} on the
Keck~I 10m telescope and the dual-channel Kast spectrometer \citep{kast} on the 3m Shane telescope
at Lick Observatory (see Table~\ref{table:obs}).

Two long slit spectra were taken using LRIS with a 1.0$''$ slit width,
  D560 dichroic, 600/4000 grism, and 600/7500 grating tilted to an angle of 28.15$^\circ$. 
This gave a red wavelength coverage of roughly 5600 to 8200\AA, a dispersion of
0.63 \AA/pixel, and FWHM~$\approx 160\  \mkms$.
In the blue we had coverage of 3100 to 5600 \AA  \,  
with FWHM~$\approx 240 \  \mkms$. 
Subsequently we took further LRIS observations with the grating angle 
tilted to 29.15 giving us coverage of 6300-9500 \AA \ to cover
H$\alpha$ and [\ion{N}{2}] emission of our z=0.37 system.  
We also took a higher resolution spectrum covering the [\ion{O}{3}] emission lines to
better characterize the gas kinematics.
We used the
0.7$''$ long slit, D560 dichroic, and 1200/7500 grating blazed to an angle of 40.15.
 This gave a dispersion of 0.4\,\AA/pix$^{-1}$, 
FWHM$~\approx 65 \ \mkms$, and wavelength 
coverage of roughly $6100-7000$\AA.
The FWHM for each of these setups was 
estimated by measuring the parameters of Gaussian models 
fit to sky lines in the science exposures.

All of the above spectra were taken with the long slit aligned 
along the line 
connecting the two sources (so that both the quasar and
 companion galaxy were within the slit). Another spectrum was taken using the Lick
 Observatory's Shane 3-meter with the Kast spectrograph. 
This spectrum was taken with a position angle (PA) 
perpendicular to the line connecting the quasar/galaxy pair, 
roughly halfway in between (such that neither the quasar nor the
companion were in the slit).
This spectrum was taken with the d55 dichroic and 600/7500 grating. 
This grating was tilted to span a wavelength range of 
approximately $5500-8300$\AA\  with a dispersion of 2.37 \AA/pixel.

Additionally we imaged the system with the LRIS camera for one 60, one 200, and two 230 second exposures in the $B$ band as well as one 90, and three 200 second exposures in the $R$ band (2008-10-03). The seeing was $\approx 0.64''$ FHWM.

We reduced the spectra using the \textsc{Low-Redux}\footnote{http://www.ucolick.org/$\sim$xavier/LowRedux/} 
pipeline developed by J. Hennawi, S. Burles, D. Schlegel, and J. X. Prochaska. We implemented this program to perform the following procedures: 
(1) process the flats, 
(2) use arcs to make a 2-dimensional wavelength image, 
(3) make a slit profile, 
(4) process the images, 
(5) identify objects in the slit and trace them, 
(6) perform sky subtraction, 
(7) extract the spectra, 
(8) correct for flexure of the spectrograph using sky lines, and 
(9) create a sensitivity function from a standard star. 
Slight modifications to the standard pipeline were necessary to
perform sky subtraction due to the extended emission and the proximity
of the galaxy to the quasar. Specifically, we turned off local sky
subtraction after verifying that a 
global fit provided a good estimate of the sky background
in the region of interest. 
Next, we coadded the images using \textsc{Long-Coadd2d}, 
a code developed by Robert da Silva and Michele Fumagalli, removing cosmic rays through comparisons of the separate exposures.

 The atmosphere and instrument response influence 
the intensity and shape of the recorded spectrum.
For point sources, the standard procedure to convert 
the observed spectrum into astronomical flux units is to 
construct a sensitivity function from
the spectrum of a spectrophotometric standard star observed
with the same instrumental configuration. 
This is done in 1-dimension, meaning that a 1-dimensional spectrum 
is extracted from the 2D output spectrum from the 
instrument and is then compared with a 1D model. 
We performed this procedure and applied the resulting sensitivity
function to the objects extracted in the slit. 

Unfortunately the region near the dichroic ($\lambda \approx 5400-5600$\AA) 
was particularly difficult to calibrate for the 2008-10-03 observations. In particular,
we find that the fluxing resulted in a galaxy spectrum that had an unphysical behavior
near the dichroic. 
We correct for this through comparison of our Keck/LRIS spectrum of the QSO and the SDSS
spectrum of the same QSO. Although the absolute and relative fluxes of quasars are known to vary,
the color variation is generally small for 
rest wavelengths greater than 2500 \AA\ \citep{wilhite05}. Avoiding issues that may arise from 
Fe emission lines we only 
sought to fit the general shape of the continuum across the boundary.
In order to account for any discrepancies between ours and the SDSS
fluxing, we matched the spectra by allowing our Keck/LRIS spectrum to be multiplied by a polynomial in
wavelength. To find the best such 2nd order polynomial we used 
the singular valued decomposition method. 
Once the overall shapes of the spectra was matched using the above procedure, we 
fit a 3rd degree polynomial to each side of the dichroic
for both the SDSS and matched LRIS 2008-10-03 spectra. 
The ratio of those polynomial fits was used to
correct the LRIS spectra. Thus the spectrum correction forces the LRIS spectrum of the
QSO to have the same shape in the region around the dichroic as the SDSS spectrum. This same correction 
was then applied to the companion galaxy spectrum.

The final product from the above procedure was a sensitivity function that
converts electrons per second per spectral pixel into astronomical flux units.
This allows us to calibrate the extracted, 1D spectrum of any object
in the slit. Our data, however, features extended emission that 
spans many arcseconds spatially.  Therefore we also require a 2-dimensional flux calibration 
that converts electrons per second per detector pixel into surface brightness units. 
To accomplish this task, we needed to apply the 1-dimensional fluxing (which has a 
mapping of pixel to sensitivity that is accurate at the center of the slit), to the entire
 2D spectrum. Assuming that the appropriate correction factor is only a function of 
 wavelength (specifically that there was a 1-to-1 mapping of wavelength to sensitivity) 
 we can apply this correction to the entire CCD using the 2-dimensional wavelength 
 image constructed with \textsc{Low-Redux} from the arc-lamp images. Interpolation of our
  sensitivity function evaluated at every pixel's wavelength gives a correction factor which we then apply to the spectra.

Since we had taken long slit data,  each of our spectra had portions
of the CCD illuminated only by the sky. We used these to test the
1-to-1 assumption described above by measuring sigma-clipped medians
along the spectral direction across spatial regions. The spatial
variation in sensitivity along the slit was found to be at most a 1\%
effect.

\section{J2049-0012: Observational Characteristics}\label{sec:ovr} 

This section provides a qualitative overview of the properties of the
J2049-0012 system. More quantitative analysis is presented in the
sections that follow. 


\begin{deluxetable}{ccccc}
\tablewidth{0pc}
\tablecaption{Compiled Photometric Measurements}
\tabletypesize{\footnotesize}
\tablehead{\colhead{} & \colhead{Quasar} & \colhead{Galaxy} & \colhead{Extinction\tablenotemark{a}} & \colhead{Ref}
\\ & (mag)&(mag) &(mag) & }
\startdata
$FUV$ & $18.54\pm 0.108$ & \nodata & 0.46 & \tablenotemark{1}\\
$NUV$ & $18.44\pm 0.063$ & \nodata & 0.49 & \tablenotemark{1}\\
$u$ & $18.05\pm 0.01$ & $21.42\pm 0.28$ &  0.46 & \tablenotemark{2}\\
$g$ & $17.89\pm 0.01$ & $20.93\pm 0.05$ & 0.34 & \tablenotemark{2}\\
$r$ & $17.79\pm 0.01$ & $19.87\pm 0.03$ & 0.25 & \tablenotemark{2}\\
$i$ & $17.75\pm 0.01$ & $19.62\pm 0.04$ & 0.19 & \tablenotemark{2}\\
$z$ & $17.10\pm 0.01$ & $19.46\pm 0.11$ & 0.13 & \tablenotemark{2}\\
$J$ & $16.51\pm 0.142$ &\nodata & $0.08$ & \tablenotemark{3}\\
$H$ & $15.74\pm 0.150$ &\nodata & $0.05$& \tablenotemark{3}\\
$K$ & $14.62\pm 0.101$ &\nodata & $0.03$ & \tablenotemark{3}\\
\tablenotetext{a}{Calculated using $A_V=0.292$ from \cite{schlegel} and assuming the
dust extinction law from \cite{cardelli_dust}. All magnitudes listed
have not been extinction corrected.}
\tablenotetext{1}{\cite{tram07}}
\tablenotetext{2}{\cite{sdssdr7}}
\tablenotetext{3}{\cite{2mass}}

\enddata
\label{table:photometry}
\end{deluxetable}

\begin{figure}
\plotone{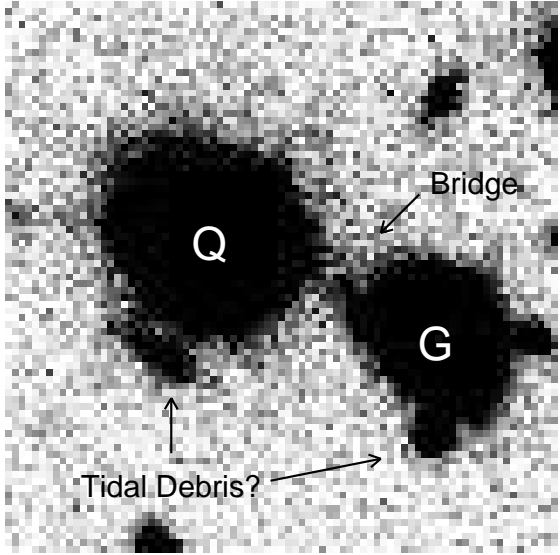}
\caption{ R-band image (which covers the redshifted [\ion{O}{3}] and
  H$\beta$ lines) zoomed in on the position of the galaxy/quasar
  pair. The image is 17$''$ on a side and is oriented so that N is up and E is left.
One can discern the bridge connecting the two galaxies as well as some possible tidal debris. The quasar (Q) and companion galaxy (G) are separated by 8$''$, corresponding to  $38 h_{72}^{-1}$ kpc at $z=0.3693$. }
\label{fig:keckimg}
\end{figure}

\subsection{Imaging}\label{sec:imaging}
This quasar/galaxy pair was selected from
the SDSS Data Release 5 \citep{sdssdr5} and GALEX DR3 \citep{galex} catalogs 
as a candidate system for an HST/COS survey 
designed to study the gas surrounding $z \sim 0.2$
galaxies with absorption-line spectra of UV bright, background quasars
(Cycle 17, ID=11598; PI: Tumlinson).
The quasar photometry and redshift ($z_{\rm q} = 0.3693$) were known
from SDSS observations while the galaxy's SDSS photometry yields a 
photometric redshift $z_{\rm phot} = 0.28 \pm 0.09$ \citep{sdssphotz}, suggesting it lay foreground
to the quasar.  Fig.~\ref{fig:keckimg} presents the R band imaging from Keck/LRIS.
The quasar (\qsonm) is the bright point source at the center of the image
and the galaxy (\jgalx) is offset by 8$''$ 
(38 $h_{72}^{-1}$ projected kpc) to the SW (lower right). 

\begin{figure}
  \epsscale{1.2}
  \plotone{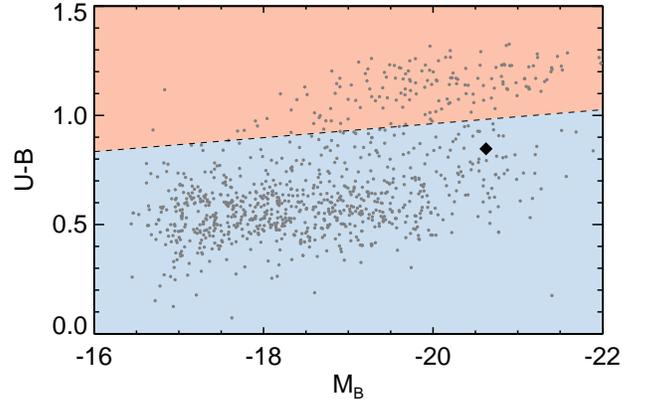}
  \caption{A rest-frame color-magnitude diagram of galaxies in 
    the redshift slice $0.3<z<0.4$ from the DEEP2 survey
    \citep[grey points;][]{deep3}. The companion galaxy of the
    J2049-0012 system (z=0.37) is represented by a black
    diamond. It appears in the `green valley'' between the main
    loci of the blue
    cloud and red sequences. The dotted line marks the separation
    between the red sequence and blue cloud as established by
    \cite{willmer2006}. 	
  } 
  \label{fig:cmd}
  \epsscale{1}
\end{figure}

The companion galaxy as seen in an SDSS finding
chart\footnote{http://cas.sdss.org/dr7/en/tools/chart/chart.asp?ra=312.48392\&dec=-0.20136} is a peculiar green,
 uncommon for galaxies observed in the SDSS.
This is illustrated in 
Fig.~\ref{fig:cmd} which compares the restframe color and absolute magnitude 
of this galaxy to a sample of DEEP2 galaxies 
at $0.3<z< 0.4$ \citep{deep1, deep2, deep3, willmer2006}.  
We find that the color of \jgalx\ places it between the two primary
populations of red and blue galaxies. 
Such galaxies are commonly described as lying within the `green valley',
a region that possibly marks the transition from blue star-forming galaxies
to red, quiescent systems \citep{faber07}.  The processes that drive galaxies
into the green valley are not well established. Scenarios include
merger-induced star bursts 
where the tidal forces 
the galaxies exert on each other funnel gas towards their centers
\citep{quint04},  which in turn 
fuels a massive starburst and ultimately a bright quasar phase \citep{hop08framework1}. 
Such a model predicts that the green valley region should coincide
 with AGN activity, which is 
supported by recent observational evidence \citep[e.g.][]{schaw09agngreenval,
 nandra07}.   However, the level of AGN activity in the companion galaxy in our quasar/galaxy pair 
 is unclear.
The 1d extraction of our galaxy shows faint [\ion{Ne}{5}]$\lambda\lambda3426,3346$ emission 
(a commonly used indicator of AGN activity), but even
this emission may be the result of 
photoionization by the neighboring quasar.

Additionally, such a model may also predict that objects in the `green valley' 
have transitional stellar populations such as those of post-starburst
galaxies. This is supported by observational studies
\citep[e.g.,][]{vergani, kocevski}. These galaxies have their
light dominated by a stellar population that has aged past the point
where its stellar light is dominated by O and B stars. Examination of
our spectra ($\S$\ref{sec:1dspectra}), suggests that our galaxy may have such
a stellar population with some ongoing star formation.  

An alternative explanation for the companion galaxy's location in the `green valley' may
be that it is a blue star forming galaxy seen nearly edge on such that dust makes it appear redder than it would otherwise \citep{bram09}.
As noted in the following section, there is evidence of 
strong rotation in the galaxy which may suggest that 
the galaxy is inclined, however there is little evidence for strong
dust absorption from the observed H$\alpha$/H$\beta$ ratio.

\begin{figure*}
\epsscale{1}
\plotone{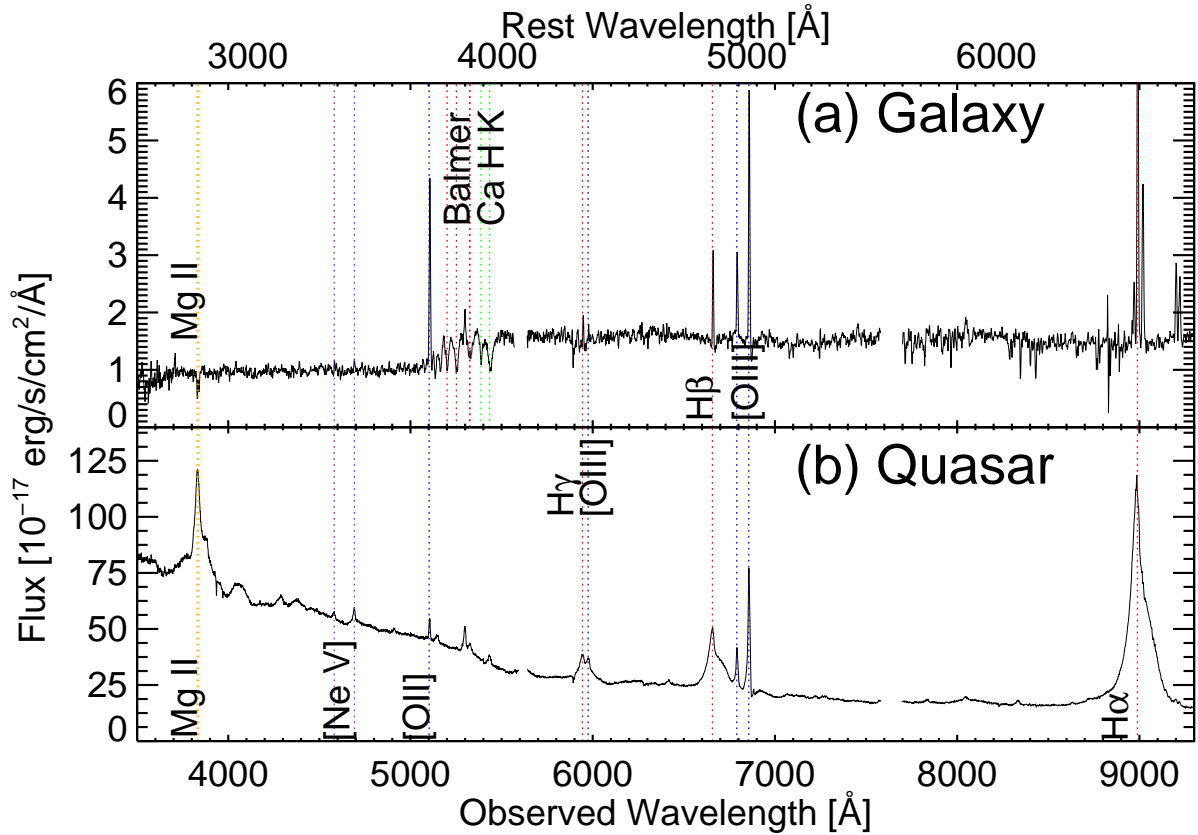}
\caption{One dimensional Keck/LRIS spectra of the galaxy and quasar 
comprising our interacting pair (the J2049$-$0012 system). Note the
strong Balmer series absorption 
lines in the galaxy spectrum; these are suggestive of a starbust population. Also note
that the quasar and galaxy are at nearly identical redshift. The quasar is at $z_{\rm q} = 0.3691 \pm 0.001$ 
and the companion galaxy is offset by only $160\pm20$ \kms. As a clarification, we note that this
quasar velocity is not the systemic velocity of its host galaxy. We find the velocity difference between the
two galaxies to be $30\pm 30$ km/s.
The strongest absorption and emisson lines are labelled by their ion.
Fig.~\ref{fig:ssp} shows an enlarged view of the post-starburst features along with models
used to constrain the age of the stellar population.
}
\label{fig:1dspec}
\epsscale{1}
\end{figure*}

Morphological analysis of the 
two objects is difficult. Firstly, the quasar (combined with its host galaxy) is consistent with a point source. The
companion galaxy, meanwhile, is resolved with only a few spatial elements 
(observed to have FWHM of 1.86'' [9.5 kpc] 
when our seeing was $\approx 0.64''$) and is also consistent 
with having an axis ratio of 1. 
Following \cite{lotz2008} and \cite{conselice2003}, we measure a
concentration for the companion galaxy of
$C=5\log_{10} \left(\frac{r_{80}}{r_{20}}\right)$
where $r_{80}$ is the radius containing 80 
percent of the total light, $r_{20}$ is similarly the
radius containing 20 percent of the total light, and we have defined 
the total light to be the light within 1.5 times the Petrosian radius.
We find that the galaxy has a concentration of 2.61 classifying it as a spiral
(the standard dividing line is spirals have $C<3$).
However, we note that since $r_{20}$ is within one seeing element, our measured concentration
is only a lower limit and therefore the profile may actually be more consistent with an elliptical or 
bulge dominated system with a higher concentration.
Around both galaxies there is some hint
in the deeper R-band exposure of tidal debris,
but since those regions did not fall into our slit we have no spectroscopic information about them and it is difficult to determine 
if this emission is truly associated with the galaxy or only close in 
projection (see Fig.~\ref{fig:keckimg}).

Table~\ref{table:photometry} lists photometry for the
two objects compiled from a variety of surveys, spanning
UV to near-IR frequencies.

Examination of the photometric redshifts of nearby galaxies reveals no obvious 
overdensity at the quasar's redshift. We examined this by looking at the number of objects
within 1 Mpc projected distance from the QSO with consistent photometric redshifts ($1.5 \sigma$) measured for
 1000 random positions in the SDSS footprint that contained at least a single galaxy (at any redshift). The quasar 
 appears in the low end of this distribution and hence there is no evidence for the quasar being in a rich and large cluster.

\begin{figure*}[!ht]
\epsscale{1.0}
\plotone{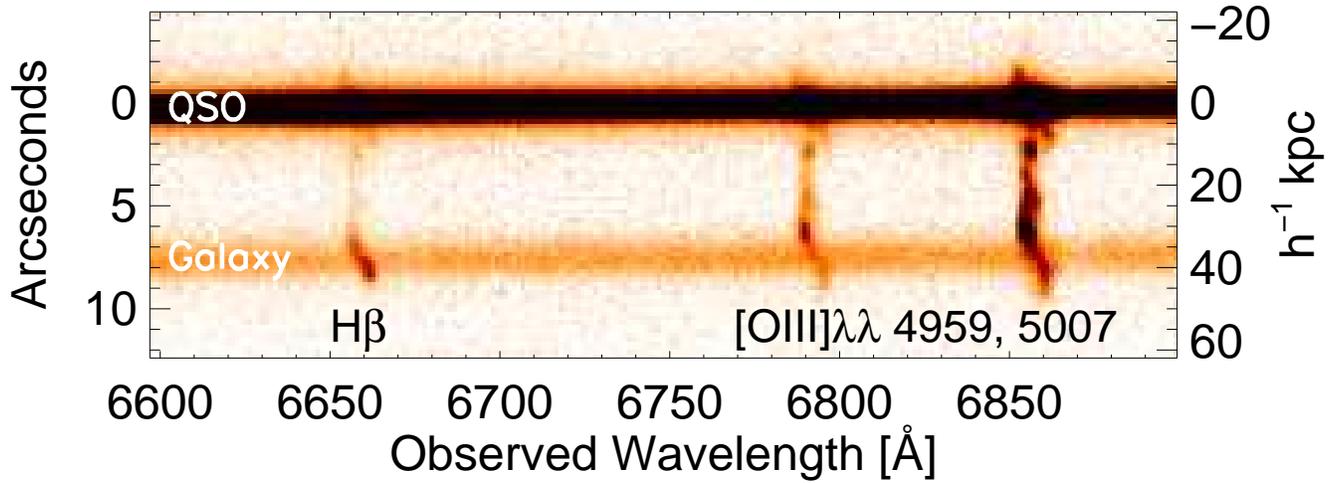}
\caption{ 2D discovery spectrum of the J2049-0012 quasar/galaxy 
system centered near H$\beta$ and [\ion{O}{3}] emission made
using the Keck/LRIS 600/7500 grating. 
The continuum light of the quasar (top) and galaxy (bottom) are the
bright horizontal bands in the image. 
The bright quasar spectrum is centered at the 0 kpc (0$''$) spatial
position and saturates the image in this stretch.  The image covers,
from left to right,  
the H$\beta$, [\ion{O}{3}]$\lambda4959$, and
[\ion{O}{3}]$\lambda5007$ transitions corresponding to $z=0.37$.  These
emission lines are detected for the quasar, its host galaxy, and the
companion galaxy as well as a `bridge' of emission that connects the two objects.
The emission lines are not detected beyond the spatial slice presented here. 
}
\label{fig:2dspec}
\epsscale{1}
\end{figure*}

\subsection{1D Spectra of the QSO Host and Companion}\label{sec:1dspectra}
The discovery spectrum was taken with the original goal of confirming this system as a projected background/foreground, quasar/galaxy system separated by
a cosmological distance.
We observed the objects using the Keck/LRIS spectrometer with its 1$''$ long slit
aligned across the pair.  Fig.~\ref{fig:1dspec} shows the fluxed 1D spectra.
The quasar shows strong and broad H$\beta$ and \ion{Mg}{2} emission lines
confirming its SDSS classification and redshift.  A Gaussian fit to the 
[\ion{O}{3}] emission lines gives $z_{\rm q} = 0.3691 \pm 0.001$ 
which is within the errors of the SDSS measured redshift of  $0.3693\pm0.0015$.
The broad emission lines appear unusually asymmetric.

The companion galaxy's photometric redshift placed it well in front of
the quasar, however the spectrum reveals the galaxy has strong
emission and absorption features at nearly the identical redshift of
the quasar with only a $\delta v \approx 150$ km/s second
offset\footnote{ This velocity difference, however, does not precisely
  reflect the velocity difference between the two galaxies which is discussed in
  $\S$\ref{sec:kinematics}. }
(Fig.~\ref{fig:1dspec}). 
The spectrum is notable for strong
emission lines characteristic of photoionization together with strong Balmer series absorption
lines that indicate a moderate-aged stellar population ($t_{\rm age} \gtrsim 100$ Myr). 
The absorption lines may  suggest that the galaxy has recently undergone a
`burst' of star-formation that may now be fading. 
After O and B stars die after $\sim$100 Myr, A stars remain a significant contributor
to the stellar light until their death approximately 1 Gyr later. Type A stars provide characterisitic
Balmer series absorption lines that make them easy to distinguish from other stellar types.
 If there is an older stellar population ($t>1$ Gyr) contributing to the galaxy light, the
 K giants will show strong Ca H+K absorption lines \citep{dress83}.
The galactic spectral type that is characterized by both Balmer absorption lines and Ca H+K lines 
and with no emission lines
is commonly referred to as  K+A post-starburst, a relatively rare
phase for galaxies especially at low redshift \citep{wild09}. 
These indicators suggest
that the galaxy's star formation rate is fading and/or has recently shut off.
A common scenario used to explain such a signature is a passively 
evolving stellar population that underwent a burst of star formation that 
has recently subsided \citep{dress83}.
Such bursts may be caused by tidal interactions 
in a merging galaxy system \citep[Snyder et al. in prep.;][]{mihos96}. This applicability of this interpretation for 
our system is 
further discussed in $\S$\ref{sec:stellarpops}.

Because the galaxies are at nearly the same redshift, there is little information that can be used to determine which galaxy is foreground.
\ion{Mg}{2} absorption is clearly visible in the companion galaxy's spectrum, while none is evident in the spectrum of the quasar (See Fig.~\ref{fig:1dspec}). 
This could indicate that the companion galaxy is background to the
quasar host galaxy \citep[e.g.][]{rubin2010}. However, the similar velocities of the two galaxies 
precludes one from unambiguously associating the gas with either the quasar host or the companion galaxy. 

\subsection{2D Spectrum}\label{sec:2dspectra}
In Fig.~\ref{fig:2dspec}, we present a slice of the co-added, 2D discovery spectrum
of the system centered near the observed H$\beta$ and [\ion{O}{3}] emission lines.
 There are two sources with visible 
continua which appear as horizontal lines in this 2-dimensional spectrum: the quasar and the companion galaxy.
  The [\ion{O}{3}] and H$\beta$ lines 
 exhibit faint emission oriented at 
$\approx 20^\circ$ counter clockwise from vertical that extends $\approx 2''$ (10 \hkpc) away 
from the spatial location of the quasar in directions both towards and away from the companion galaxy. 
We associate this emission 
with the underlying host galaxy of the quasar. The 
weaker continuum source located at impact parameter $\rho \approx 8''$ (38 \hkpc) and its 
associated emission lines are from the companion galaxy.
The galaxy emission shows obvious signatures of rotation; the [\ion{O}{3}]
emission spans $\sim$5 \AA\ corresponding to $\Delta v = c \Delta \lambda/\lambda 
\approx  300 \ \mkms$
which implies a rotation speed $v_{\rm rot} > 100 \mkms$. Interestingly, the 
galaxy's emission properties vary as a function of spatial distance from the quasar: H$\beta$ emission is stronger 
on the side away from the
quasar than the side toward it, while [\ion{O}{3}] shows the opposite trend.

Of greatest interest to this work is the detection of 
a `bridge' of emission that connects the quasar and galaxy.
The bridge is strongest in [\ion{O}{3}] but H$\beta$ and other lines shown 
in Fig.~\ref{fig:lineprofs} 
are also positively detected.
 It appears that the bridge lies roughly along a line connecting the spatial and velocity
 centers of the two galaxies.
Fig.~\ref{fig:2dspec} also reveals that along the line connecting
the quasar and galaxy, the emission is confined to the space between them.
In this respect, the emission differs from the EELRs commonly observed
around quasars \citep[e.g.][]{fu09}, i.e.,  
it appears to be directly related with the gas in this interacting system. 
To explore this point further,
we obtained an optical spectrum with the Kast spectrometer on the
3m Shane telescope at Lick Observatory with the long slit centered on
the bridge but oriented perpendicular at a PA perpendicular to that of our Keck/LRIS spectrum.
By fitting a gaussian to the Kast spectrum, we found its FWHM=2.0$''$, which
was consistent with our seeing limit for the night. Therefore we 
constrain the spatial extent of the bridge to be 
$< 10.2 \ \mhkpc$ perpendicular to the line connecting the quasar and companion galaxy.

\subsection{Summary}\label{sec:obs_sum}

\begin{figure}
\plotone{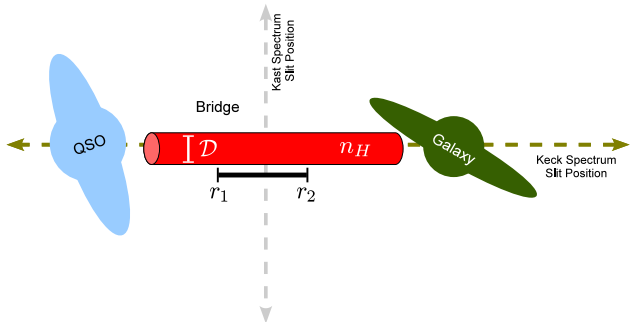}
\caption{A cartoon depiction for the assumed geometry of the \qsonm\ system. 
We assume that the bridge forms a cylinder of diameter $\mathcal{D}$ connecting the quasar host galaxy to the companion galaxy.
$r_1$ and $r_2$ are radial distances from the quasar. The volume density of hydrogen is assumed to be a constant $n_H$. Also noted are the
approximate slit positions of both the Keck/LRIS and Shane/Kast
spectra. Using this simple model, we can 
estimate properties of the bridge including its volume and column
densities and its mass.}
\label{fig:cartoon}
\end{figure}

Fig.~\ref{fig:cartoon} presents a cartoon
of the inferred layout of the system.
Taken altogether, the observations of this system suggest the
following scenario: we are observing a pair of merging galaxies that have undergone an interaction
 which has tidally stripped material from at least one of the
 galaxies. This is suggested by the nearly identical redshifts of the galaxies and
 the presence of the bridge connecting the two galaxies.
Such material is a standard prediction in theoretical models of gas-rich
galaxy encounters \citep{toomre72, mihos96} as well as a commonly observed phenomenon 
in nearby, interacting galaxies \citep[e.g.][]{arp,roguesgallery}.
The ongoing merger could have triggered a burst of star-formation in one
(or both) of the galaxies and AGN activity in the galaxy now observed to be in a quasar phase. The companion galaxy appears
to be in the `green valley' and has a possibly fading starburst stellar population.
This population is consistent with a starburst that was caused during
the first passage of a galaxy interaction and may indicate that the
galaxy is likely undergoing a substantial change in its stellar light
properties. The fact that the higher ionization state lines are
brighter closer to the quasar ($\S$\ref{sec:lineratios}) suggests that  
the quasar is shining on the tidal material,
photoionizing the majority of that gas.  If this hypothesis is correct, then the full toolbox 
of low density astrophysics allows us to interpret  observations of the bridge's 
emission lines to give insights into the tidal 
material's dynamics, column density, mass, temperature, metallicity, and volume density.   
In the following sections we test this physical interpretation of the system and its implications. 

\section{Emission Line Measurements}
\begin{figure}
\epsscale{0.9}
\plotone{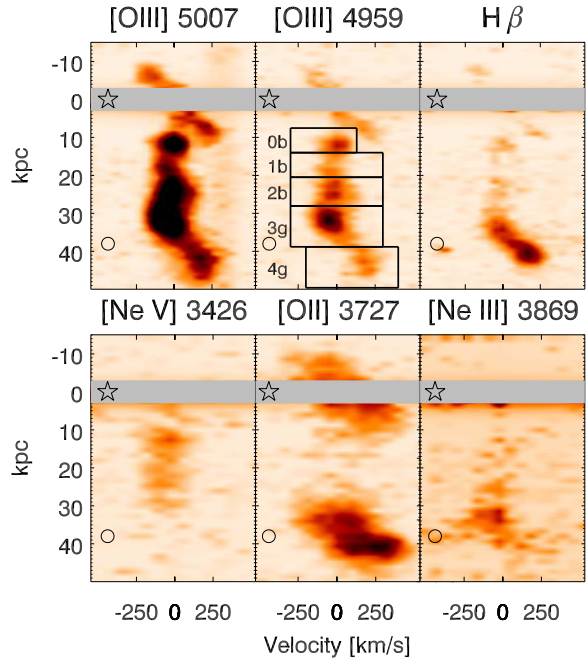}
\caption{Portions of the discovery spectrum taken with Keck/LRIS on 2008-10-03
after continuum subtraction. The star denotes the spatial position of
the quasar and the open circle marks the position of the companion
galaxy. A grey bar has been placed over the core of the quasar to aid
the eye. These images have been interpolated to be on the same spatial
and velocity grid. The [\ion{O}{3}]$\lambda 4959$ panel shows the
spatial boxes used throughout the paper which separates regions
in each of the bridge (0b, 1b,2b) and the companion galaxy (3g,4g).}
\label{fig:lineprofs}
\end{figure}

The flux measurements and line ratios constrain
properties of the gas and galaxies involved in this system. The kinematics 
provide estimates for the masses of the galaxies and give information about the
merger stage and geometry. We present the emission line fluxes in Table~\ref{table:linefluxes}. 
Inspection of Fig.~\ref{fig:lineprofs} reveals that the emission line fluxes
and the line ratios vary along the slit hence
 we have broken the bridge into three bins (0b, 1b, and 2b) and
 the galaxy into two equally sized bins (3g and 4g) as labelled in
Fig.~\ref{fig:lineprofs}. The bounds of the regions noted as
``bridge'' and ``galaxy'' are determined by eye
through inspection of the kinematic profiles. In the higher resolution spectrum, 
it is clear that the bridge has distinct kinematic characteristics from either galaxy. 

\begin{deluxetable*}{ccccc}\label{table:fluxes}
\tablecaption{Emission Line Measurements}
\tablehead{\colhead{}  & \colhead{Region\tablenotemark{a}}&\colhead{Flux} & \colhead{Dereddened Flux\tablenotemark{b}} & \colhead{$\frac{Flux}{F_{5007}}$ \tablenotemark{c}} \\ 
\colhead{}& \colhead{} &\colhead{$10^{-17} \, \mbox{erg/s/cm}^2$}& \colhead{$10^{-17} \, \mbox{erg/s/cm}^2$}&  \colhead{} }
\startdata

 [\ion{N}{2}]$\lambda$6585	 & 0b & $<$0.87\tablenotemark{d}&$<$0.99&$<$0.07\\
  & 1b & 1.34$\pm$0.38 & 1.52$\pm$0.43 & 0.12$\pm$0.03\\
  & 2b & $<$1.29&$<$1.46&$<$0.05\\
  & 3g & 11.68$\pm$0.60 & 13.28$\pm$0.68 & 0.27$\pm$0.01\\
  & 4g & 13.42$\pm$0.67 & 15.26$\pm$0.76 & 0.75$\pm$0.04\\
\hline\\[.1 mm]
 H$\alpha$ & 0b & 4.89$\pm$0.43 & 5.56$\pm$0.49 & 0.38$\pm$0.04\\
  & 1b & 3.53$\pm$0.50 & 4.02$\pm$0.57 & 0.31$\pm$0.04\\
  & 2b & 5.19$\pm$0.56 & 5.90$\pm$0.64 & 0.22$\pm$0.02\\
  & 3g & 32.51$\pm$0.77 & 36.98$\pm$0.88 & 0.77$\pm$0.02\\
  & 4g & 38.18$\pm$0.79 & 43.44$\pm$0.89 & 2.14$\pm$0.08\\
\hline\\[.1 mm]
 [\ion{N}{2}]$\lambda$6549	 & 0b & $<$0.92&$<$1.04&$<$0.07\\
  & 1b & $<$1.18&$<$1.35&$<$0.10\\
  & 2b & $<$1.32&$<$1.50&$<$0.06\\
  & 3g & 4.48$\pm$0.60 & 5.10$\pm$0.68 & 0.11$\pm$0.01\\
  & 4g & 4.95$\pm$0.64 & 5.64$\pm$0.73 & 0.28$\pm$0.04\\
\hline\\[.1 mm]
 [\ion{O}{3}]$\lambda$5007 & 0b & 11.87$\pm$0.30 & 14.62$\pm$0.37 & \nodata\\
  & 1b & 10.65$\pm$0.35 & 13.11$\pm$0.44 & \nodata\\
  & 2b & 21.85$\pm$0.41 & 26.90$\pm$0.51 & \nodata\\
  & 3g & 39.24$\pm$0.55 & 48.30$\pm$0.67 & \nodata\\
  & 4g & 16.52$\pm$0.53 & 20.34$\pm$0.65 & \nodata\\
\hline\\[.1 mm]
 [\ion{O}{3}]$\lambda$4959 & 0b & 4.27$\pm$0.26 & 5.27$\pm$0.32 & 0.36$\pm$0.02\\
  & 1b & 3.70$\pm$0.31 & 4.57$\pm$0.39 & 0.35$\pm$0.03\\
  & 2b & 7.41$\pm$0.36 & 9.14$\pm$0.44 & 0.34$\pm$0.02\\
  & 3g & 12.93$\pm$0.48 & 15.96$\pm$0.60 & 0.33$\pm$0.01\\
  & 4g & 4.77$\pm$0.46 & 5.89$\pm$0.57 & 0.29$\pm$0.03\\
\hline\\[.1 mm]
 H$\beta$ & 0b & 1.30$\pm$0.25 & 1.62$\pm$0.32 & 0.11$\pm$0.02\\
  & 1b & 1.72$\pm$0.31 & 2.14$\pm$0.38 & 0.16$\pm$0.03\\
  & 2b & 2.40$\pm$0.34 & 2.97$\pm$0.43 & 0.11$\pm$0.02\\
  & 3g & 9.51$\pm$0.47 & 11.81$\pm$0.58 & 0.24$\pm$0.01\\
  & 4g & 8.18$\pm$0.46 & 10.15$\pm$0.57 & 0.50$\pm$0.03\\
\hline\\[.1 mm]
 [\ion{Ne}{3}]$\lambda$3869	 & 0b & 0.92$\pm$0.22 & 1.22$\pm$0.29 & 0.08$\pm$0.02\\
  & 1b & 1.05$\pm$0.25 & 1.39$\pm$0.33 & 0.11$\pm$0.03\\
  & 2b & 1.68$\pm$0.28 & 2.22$\pm$0.37 & 0.08$\pm$0.01\\
  & 3g & 4.08$\pm$0.37 & 5.41$\pm$0.49 & 0.11$\pm$0.01\\
  & 4g & 2.00$\pm$0.36 & 2.65$\pm$0.47 & 0.13$\pm$0.02\\
\hline\\[.1 mm]
 [\ion{O}{2}]$\lambda$3727	 & 0b & 1.18$\pm$0.21 & 1.59$\pm$0.28 & 0.11$\pm$0.02\\
  & 1b & $<$0.72&$<$0.96&$<$0.07\\
  & 2b & 2.00$\pm$0.26 & 2.69$\pm$0.35 & 0.10$\pm$0.01\\
  & 3g & 12.83$\pm$0.52 & 17.21$\pm$0.70 & 0.36$\pm$0.02\\
  & 4g & 14.30$\pm$0.49 & 19.19$\pm$0.65 & 0.94$\pm$0.04\\
\hline\\[.1 mm]
 [\ion{Ne}{5}]$\lambda$3426	 & 0b & 2.44$\pm$0.20 & 3.38$\pm$0.28 & 0.23$\pm$0.02\\
  & 1b & 2.18$\pm$0.23 & 3.03$\pm$0.32 & 0.23$\pm$0.03\\
  & 2b & 2.26$\pm$0.25 & 3.14$\pm$0.35 & 0.12$\pm$0.01\\
  & 3g & 1.24$\pm$0.32 & 1.73$\pm$0.44 & 0.04$\pm$0.01\\
  & 4g & $<$0.94&$<$1.30&$<$0.06\\
\hline\\[.1 mm]
 [\ion{Ne}{5}]$\lambda$3346	 & 0b & 0.92$\pm$0.20 & 1.28$\pm$0.28 & 0.09$\pm$0.02\\
  & 1b & 0.79$\pm$0.23 & 1.10$\pm$0.33 & 0.08$\pm$0.03\\
  & 2b & $<$0.71&$<$1.00&$<$0.04\\
  & 3g & $<$0.96&$<$1.35&$<$0.03\\
  & 4g & $<$0.96&$<$1.35&$<$0.07\\
\hline\\[.1 mm]
\enddata

\tablenotetext{a}{Region 0b=[7.9,13.8)$''$;Region 1b=[13.8, 20.7)$''$; Region
  2b=[20.7,27.6)$''$; Region 3g=[27.6, 38.63)$''$; Region
  4g=[38.63,49.67)$''$ as shown in Fig.~\ref{fig:lineprofs}.}
\tablenotetext{b}{\cite{schlegel, cardelli_dust}}
\tablenotetext{c}{Dereddened flux relative to value of [\ion{O}{3}]$\lambda 5007$ in that spatial bin.}
\tablenotetext{d}{All upper limits are at 3$\sigma$ level.}
\label{table:linefluxes}
\end{deluxetable*}

\subsection{Line Fluxes}\label{sec:linefluxes}
To properly isolate the emission line fluxes, we 
needed to first subtract the continuum light from the quasar and the galaxy. 
This required construction of an accurate model of  
the flux from each object.  We took the 2-dimensional profile 
fits from \textsc{Low-Redux} (interpolating over the emission-line 
regions) and multiplied by the 1-dimensional boxcar extraction. 
An additional complication arises from the companion galaxy's Balmer absorption 
lines. To account for these spectral features, we used the
stellar population model described in $\S$~\ref{sec:stellarpops}, to model
the absorption for the H$\alpha$ and H$\beta$ 
fluxes. This resulted in a spectrum containing
 only the emission lines (Fig.~\ref{fig:lineprofs}). 

The main contributions to the error are the 
statistical errors of the signal (Poisson errors) and the errors associated with 
our data reduction (i.e. sky subtraction and continuum subtraction). 
While the former is relatively straightforward to estimate, the latter is more difficult. 
We estimated this systematic error through analysis of our science
images that had been sky and profile subtracted as follows.
For every pixel in these images, we calculated the sigma-clipped mean and standard 
deviation in a square region with 10 pixels on a side
(after masking known cosmic rays and the emission lines themselves). 
We then have an image of the local means and an image of the local standard deviations.
 For each emission line we picked
a region from these images of means and standard deviations that is
nearby (but does not include emission) and proceeded to take their averages.
 We then 
add an additional error term in quadrature with the Poisson errors
 equal to the average of the standard deviations in this region. We also include an additional
 error associated with over/undersubtracting the profiles that is characterized by applying the entire
 profile subtraction algorithm to regions that contain no emission lines and measuring fluxes in the same velocity
 apertures in those regions.
  
Lastly,  when comparing line fluxes 
we must take care to consider the same spatial region for each of the emission lines.  
Because our observations were taken with different cameras each with a different plate scale, 
the spatial region determined by an integer number of pixels 
for one of the plate scales is not generally equal to an integer number of pixels 
for others. 
Therefore, we insure we are comparing the same region
by interpolating the spectrum onto a grid 100 times finer in
spatial and velocity resolution. 
We note that this is valid as our
platescales, seeing, and dispersion
render the data Nyquist sampled in all of the observations. We then use the spatial centroid 
of the quasar continuum light as a reference and extract along the same spatial intervals for all the lines. 
We then apply velocity extraction windows 
as shown in Fig.~\ref{fig:lineprofs}. The velocity extraction windows are broader for the [\ion{O}{2}] lines
because they form a doublet and are treated as a single line which is therefore seen to have a larger
velocity spread. The H$\alpha$ and [\ion{N}{2}] lines have slightly 
smaller velocity windows to 
prevent overlap with nearby bright skylines.

The flux measurements are presented in Table~\ref{table:linefluxes}. 
We corrected fluxes for Galactic foreground reddening using 
the Schlegel, Finkbeiner and Davis dust maps \citep[$A_V=0.292$;][]{schlegel} and a Milky Way dust extinction curve from \cite{cardelli_dust}.

\subsection{Line Ratios}\label{sec:lineratios}

\begin{figure}
\plotone{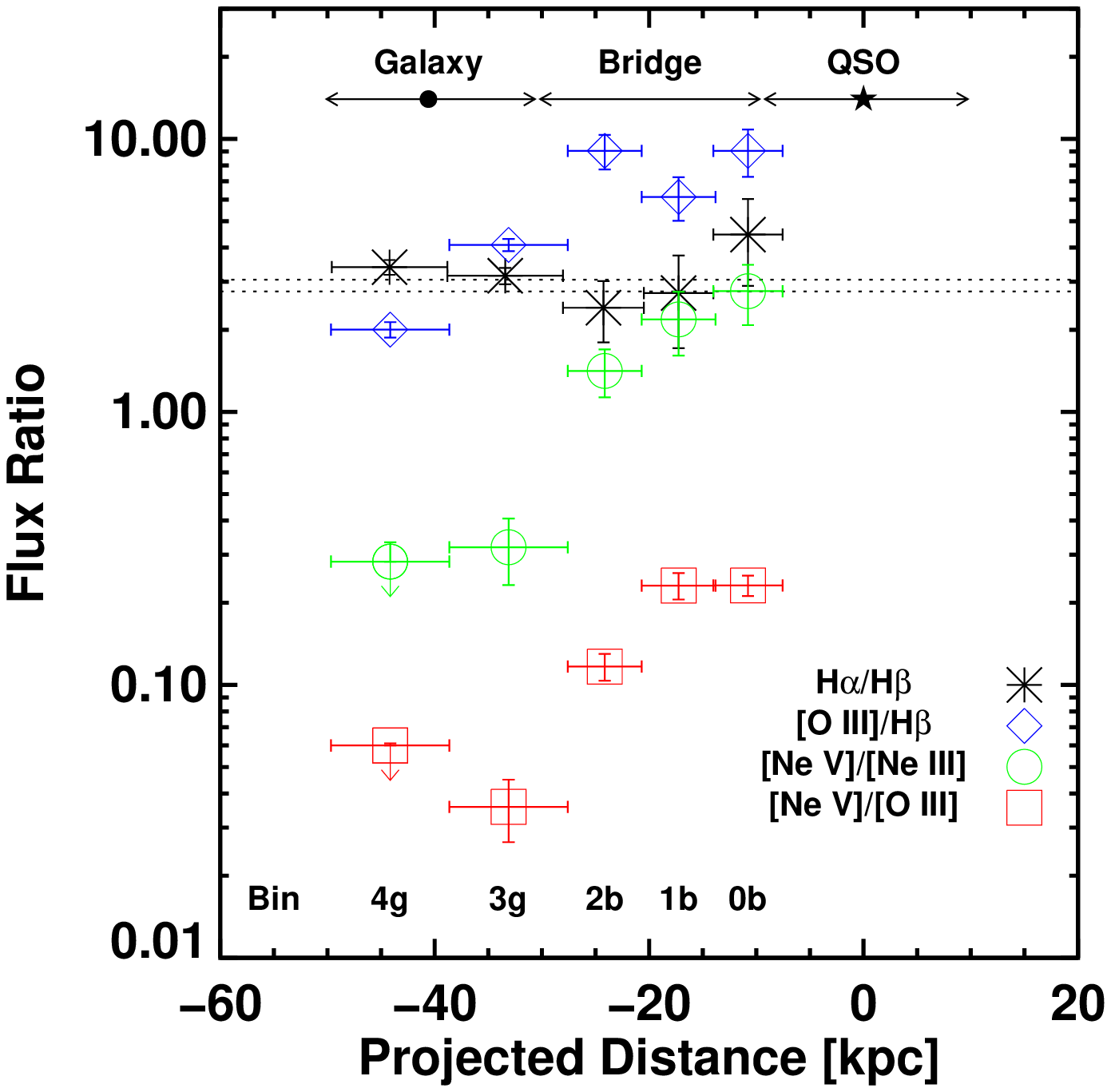}
\caption{The spatial variation of various line ratios measured for the
  \qsonm\ system. The lines used were H$\alpha$, H$\beta$, [\ion{O}{3}]$\lambda$5007,
  [\ion{Ne}{5}]$\lambda$3426, and [\ion{Ne}{3}]$\lambda$3869.  
  The dotted
  lines denote the theoretical values for the H$\alpha$/H$\beta$ ratio
  for gas at $T_e = 5,000$K (3.05) and at 20,000K (2.76) \citep{agn2}. The
  galaxy and bridge are consistent with having little intrinsic
  dust. Higher ionization lines are stronger relative to lower ones
  closer to the QSO (at spatial position 0 kpc). This trend suggests
  that it is possible the quasar is responsible for the emission
  because
  the [\ion{O}{3}]/H$\beta$, [\ion{Ne}{5}]/[\ion{Ne}{3}], and [\ion{Ne}{5}]/[\ion{O}{3}]
  ratios are all sensitive to the ionization state of the gas.
}
\label{fig:lineratios}
\end{figure}

Fig.~\ref{fig:lineratios} presents several line ratios that gauge physical characteristics of the system.
One such ratio, that of H$\alpha$ to H$\beta$ (black crosses), is a standard indicator of dust absorption. 
The dotted lines show theoretical estimates of this ratio at a range of temperatures spanning reasonable expectations for photoionized gas \citep{agn2}. 
The agreement between the observations and theoretical prediction
implies that there is little dust in the bridge and only modest
extinction within the companion galaxy.

The figure also shows how the brightness of an emission line varies with distance from the quasar. 
The apparent trend is that lines coming from higher ionization species 
reach their brightest intensities (relative to other lines) closer to the quasar while 
lower ionization species show the opposite trend. Thus the ratios [\ion{O}{3}]/H$\beta$, [\ion{Ne}{5}]/[\ion{Ne}{3}], and [\ion{Ne}{5}]/[\ion{O}{3}] are decreasing with
increasing distance from the quasar.
This suggests that the quasar is related to the observed emission (see $\S$ \ref{sec:source}).

The ratio of [\ion{Ne}{5}] to [\ion{Ne}{3}] 
constrains the ionization parameter of the gas 
(which is an indicator of the ionization state of the gas). Indeed, 
the presence of [\ion{Ne}{5}]
emission alone requires a source of high energy
photons or high temperature ($T$) gas 
(with an ionization potential exceeding
7 Rydberg, \ion{H}{2} regions will not typically produce this line).
The emission line ratios
are also important in determining the source of ionization, 
e.g. [\ion{N}{2}]/H$\alpha$ and [\ion{O}{3}]/H$\beta$
are sensitive to the ionizing source's spectrum.

\begin{figure*}
\begin{minipage}{11.0cm}
\epsscale{1.2}
\plotone{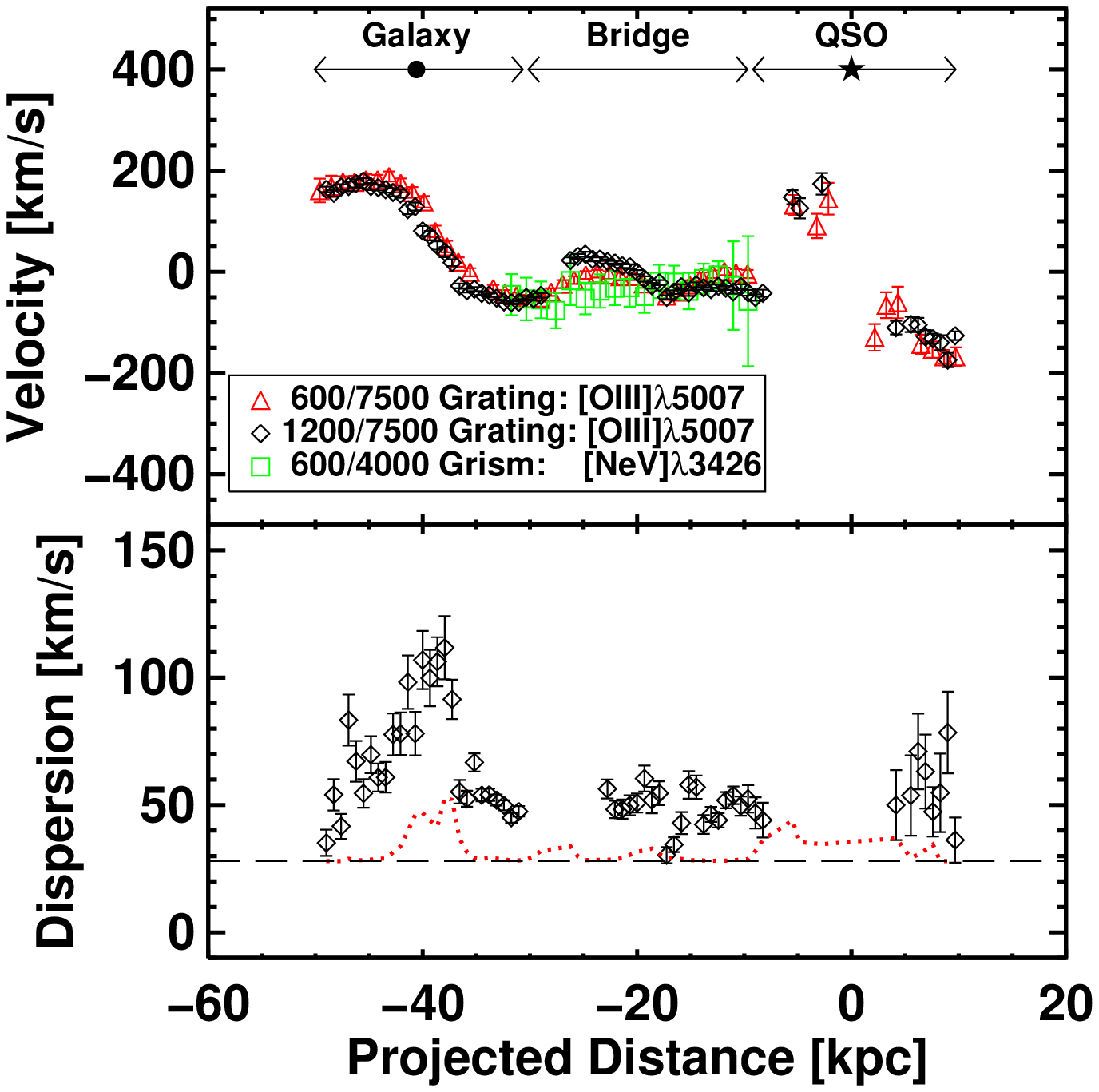}
\end{minipage}
\begin{minipage}{6.0cm}
\epsscale{1.15}
\plotone{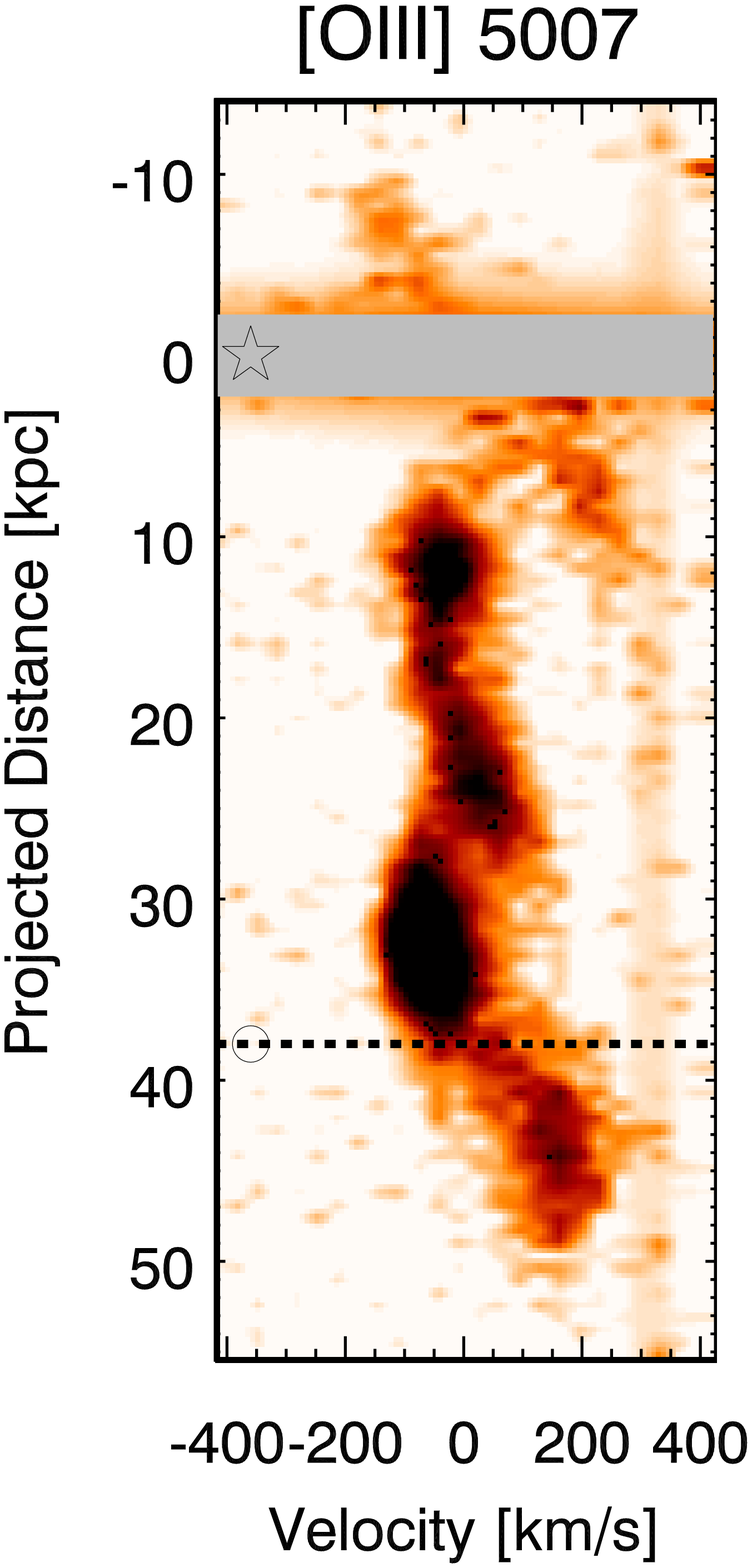}
\end{minipage}
\caption{Kinematic characteristics for several emission lines detected
  for the \qsonm\ system. The values presented on the left are derived
  from Gaussian fits to each spatial row in the 2D spectrum. 
(\emph{left top}) The velocity centroids of each fit; 0 km/s is
defined by the SDSS redshift of the QSO. 
The arrows at the top denote
the spatial extent of each region of the system. Specifically, the
star denotes the centroid of the quasar, the circle the centroid of
the galaxy and the central region is called the bridge. Note that the
[\ion{Ne}{5}] emission is kinematically coincident with the
[\ion{O}{3}] emission indicating that these ions likely arise in a
single phase of gas. (\emph{left bottom}) The velocity
dispersions ($\sigma$) measured from the higher resolution 1200/7500 grating
for the [\ion{O}{3}]$\lambda$5007 line. The dashed line is the
measured velocity dispersion of a nearby sky line. The red dotted line
represents the expected velocity dispersion as expected by the finite
spatial resolution of the spectrum (see the text). 
Spatial bins that were poorly fit by a single gaussian have been masked. 
(\emph{right}) The higher resolution (1200/7500 grating) spectrum of
the [\ion{O}{3}] emission. This spectrum shows finer detail than the
600/7500 grating and is used for comparisons with simulations and to
estimate the masses of the galaxies. One can clearly see that the bridge has
distinct kinematic properties from either galaxy, yet connects
spatially with each galaxy near their velocity centroids.
}
	\label{fig:kinematics} \label{fig:highres}
        \end{figure*}

\subsection{Kinematics}\label{sec:kinematics}

The kinematics of the emission constrain the
driving gravitational forces on the gas. They also provide
estimates for the masses of both the QSO host and its companion galaxy. Coupled with comparisons to
simulations, we can also learn about the specific merger stage of the
galaxies. 
To characterize the kinematics of the ionized gas,
we calculated the flux-weighted velocity
centroid and velocity dispersion in the [\ion{O}{3}]$\lambda 5007$ line within each spatial row of pixels. 
We choose the [\ion{O}{3}]$\lambda 5007$
line because it is the brightest observed line with the highest signal-to-noise ratio.
These measures were calculated by fitting single gaussians to all velocity pixels 
corresponding to a given spatial row of pixels.
Each fit was checked by eye and catastrophic errors related to 
non-gaussian line-profiles have been masked.  These are due to regions with
low signal-to-noise or when there are overlapping velocity components
(e.g., in spatial regions that include
both the bridge and one of the galaxies).

The dominant source of uncertainty in these kinematic measures is related
to systematic effects. Specifically, the wavelength 
solution based on the arc lamp images has a typical RMS residual of 
$\approx 0.1$\,pixels corresponding to 
a velocity of $\approx 10 \ \mkms$. The systematic velocity error due to this
effect is calculated for each line by (1) measuring the RMS of the residuals from our wavelength 
solution in units of pixels (2) multiplying the RMS by the dispersion (\AA/pix) to get the wavelength RMS 
(3) then estimating the error that is introduced at each line's observed wavelength. These errors are added
in quadrature with the fitting errors.

We find that
the kinematics of the [\ion{Ne}{5}]$\lambda 3426$ line 
trace the [\ion{O}{3}]$\lambda 5007$ emission\footnote{It is best to
  compare the [\ion{Ne}{5}]$\lambda 3426$ kinematics to those derived
  from the R600/7500 grating for [\ion{O}{3}]$\lambda 5007$ since they
 were measured with comparable resolution.}; 
it is thus very likely that these ions are co-spatial. This is a
crucial point for the determination 
of the ionization source, as it allows us to demand that the 
same ionization mechanism produce each of these
ions and their observed line ratio.

The velocity centroids and dispersions ($\sigma$) of [\ion{O}{3}]$\lambda 5007$ as measured from 
our highest resolution R1200/7500 observation are shown in Fig.~\ref{fig:kinematics}.
In our higher resolution spectrum, the kinematics of the gaseous bridge appear
distinct from the rotation observed in the quasar's host galaxy and 
the companion galaxy (See Fig.~\ref{fig:kinematics}).

The lower panel reports the measured velocity 
dispersions of the 1200/7500 measurements of the [\ion{O}{3}] $\lambda 5007$ line.  
The width of a nearby unresolved sky line is plotted as a dashed line to indicate the 
resolution limit of the spectrum. Since the velocity dispersions are higher than those of a nearby skyline 
we appear to have spectroscopically resolved
the emission of the bridge, but it is unclear if we have done so
spatially. Specifically, we emphasize that if there are spatial
gradients in the velocity field that are unresolved they have the
effect 
of artificially increasing the velocity dispersion because the gradient is smeared. 
Thus one might misinterpret spatially unresolved velocity shear 
as random motions.
To quantify this effect we created
a mock 2D spectrum with the same central velocities and brightnesses as a function of position as those measured from our data. To isolate the effects of
this spatial smearing we set the velocity dispersions of each spatial pixel equal
to that of the unresolved skyline. We then smoothed this 2D spectrum in the spatial direction according to a spatial resolution element of 0.7$''$
and measured the velocity dispersion in the same manner as for the
data. The result is represented as the red dotted curve and can be
interpreted as 
the expected effective resolution limit as a result of both finite velocity resolution and astronomical seeing. 
Since the velocity dispersion is still higher than this curve, it
appears we are also spatially resolving
the velocity dispersion (as long as the gradients are not smaller than the seeing).

The velocity of the quasar spectrum as measured from AGN emission
lines is generally a poor measure of the systemic velocity for its
host galaxy \citep[e.g.,][]{richards2002}. Therefore, we have also analyzed
the host galaxy's extended [\ion{O}{3}] emission. We use the measured 
centroids on either side of the host galaxy along with the
spatial centroid of the quasar light to interpolate the velocity 
at the spatial center of the host galaxy. We do this by fitting a robust
linear fit to the data.
Thus we measure the relative radial velocities of the quasar host galaxy and companion galaxy to
be $30\pm 30$\ \kms, with the quasar at the higher radial velocity.

\section{Physical Properties of the Interacting Pair}
In this section we present analysis detailing measured physical  properties of the
galaxies and the SMBH of the QSO involved in the merger.

\subsection{QSO Properties}\label{sec:bhprop}

One can estimate the mass of the supermassive black hole $M_{BH}$
fueling the quasar activity in several ways based on the emission line characteristics \citep{vest06, kong06}. 
By fitting SED templates one may estimate the bolometric luminosity of the QSO from the continuum luminosity in 
a particular wavelength region.
The quasar under study is in the SDSS QSO catalog of 
\cite{shen08} who used the above methods to calculate 
$L_{Bol} = 10^{45.7}$ erg/s and $M_{BH}=10^{8.7} M_{\odot}$ where each has a typical error of
approximately 0.4 dex. This places the quasar above the average for
quasars at this redshift but within approximately 1$\sigma$ of 
the distributions for both quantities. This also gives an Eddington ratio of 0.07 which falls within the normal range for quasars \citep{blqso_massfcn}.

We utilize the calibration of \cite{beyondbulge} derived using \cite{Bullock2001} to estimate the dark
matter halo mass of the QSO host galaxy, 
\begin{eqnarray}
\frac{M_{DM}}{10^{12}M_\odot}& \approx & 1.40\left(
  \frac{v_c}{200\mkms}\right)^{3.32} \;\;\; , 
\end{eqnarray}
where $v_c$ is the circular velocity of the host galaxy at the point where the rotation becomes
flat. 

Using the data from the R1200/7500 grism (see Fig.~\ref{fig:kinematics}), 
we find that emission associated with the quasar host galaxy extends
from $-174.3$ to +174.0 $\mkms$ 
(where our measurements at 
the positive velocity end are truncated when they become
cospatial with the bridge).
These velocities however are relative to
the SDSS quasar redshift and not the systemic velocity of
the quasar host. Using the methods described in $\S$\ref{sec:kinematics}, we
found the offset to be $30\pm30$ km/s. Thus we arrive at an estimate of 
$v_{c} \ge (200 \pm 30)/\sin i$ where $i$ is the galaxy inclination. 
This is a lower limit on the actual circular velocity because
(1) we may not have aligned our slit with the major axis of the galaxy and 
(2) our spectrum does not appear to extend to
the flat portion of the rotation curve. Using the 1$\sigma$ error below the measured value, we find that 
this gives an estimate 
of the host mass to be $M_{DM}>8.8\times 10^{11} \ M_\odot$.

 With IFU spectra or additional spectra at different slit angles, we could constrain
 the inclination and axis ratio of the host galaxy to better characterize the host galaxy's dynamical mass.
 Space-based imaging however would allow us to 
directly image the host's bulge and multiband imaging would allow us to better estimate its stellar mass. 
In paper II we discuss the position of the BH and host on Magorrian-like relations and details of how this and other systems like it
might lead to insights on whether a host galaxy leads/lags its SMBH.
 
\subsection{Companion Galaxy Properties}

\begin{figure}
\epsscale{1.3}
\plotone{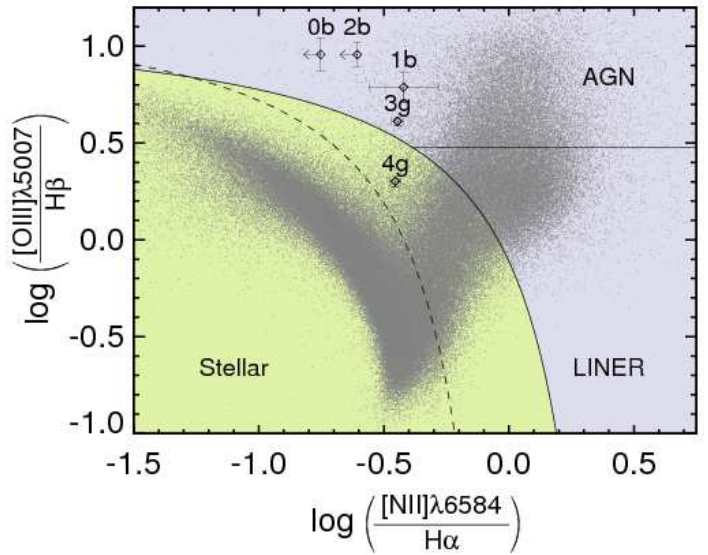}
\caption{A BPT \citep{bpt} diagram of the flux ratios observed for the bridge (0b, 1b, 2b) and companion galaxy (3g, 4g) regions as demarcated in Fig.~\ref{fig:lineprofs}. 
The points are a random selection of galaxies from SDSS to give a sense of the normal locus of objects.
The solid curve is from \cite{kewl01} and separates regions consistent with ionization from \ion{H}{2} regions from other sources of ionization. The dashed curve is a similar division found by \cite{kauffbpt}. The horizonal line segment denotes the separation between LINERS and AGN \citep{veil87}.
The regions of the bridge (0b, 1b, 2b) are consistent with being
photoionized by a hard AGN spectrum while bin 4g (located farthest
from the quasar) is more consistent with \ion{H}{2} regions. Bin 3g is
in a portion of the diagram that is inconsistent with
ionization by \ion{H}{2} regions alone and likely has an ionizing
spectrum that includes the AGN and a stellar component. }
\label{fig:bpt}
\end{figure}

\subsubsection{Star Formation Rate \& Gas Metallicity}\label{sec:metal}
The observed line ratios of H$\alpha$/[NII] and [OIII]/H$\beta$ define the
 the standard axes of the Baldwin, Phillips and Terlevich
(BPT) diagram \citep[][see Fig.~\ref{fig:bpt}]{bpt}.  
These line ratios are used because they are sensitive to the hardness of the spectrum and
the temperature of the gas (which in turn is also sensitive to the
hardness of the spectrum).
Furthermore, these are bright lines in the optical passband
that are closely spaced in wavelength such that differential reddening is
generally negligible. 
In this diagram \ion{H}{2} regions occupy a particular locus, while AGN and LINER sources
each separate into their own loci. The curved line denotes the division between emission characteristic 
of ionization from hot young stars in \ion{H}{2} regions from ionization 
from other sources \citep{kewl01}. The dashed curve is a similar division
as reported by \cite{kauffbpt}.
The horizontal line segment to the right 
of this curve at a value of $\log($[\ion{O}{3}]$\lambda 5007/H\beta)=\log(3)$ 
denotes the division between the LINER and Seyfert line ratios 
as defined by \cite{veil87}.

The galaxy spans bins 3g and 4g. While bin 3g appears strongly affected by
an AGN-like source, bin 4g is closer to the locus of gas photoionized by stars.
Under the hypothesis that the ionization in region 4g of the companion galaxy
is consistent with \ion{H}{2} regions,
we can estimate 
the star formation rate using the \cite{kennicutt98} calibrations for H$\alpha$:

\begin{eqnarray}
SFR_4(M_\odot \textnormal{  yr}^{-1})&=&7.9\times 10^{-42}L(\textnormal{H}\alpha)(\textnormal{ergs s}^{-1})^{-1} \nonumber\\
&=&1.61\pm 0.03 \ M_\odot\textnormal{  yr}^{-1} .
\end{eqnarray} 
Assuming that the [OII] emission also arises from \ion{H}{2}
regions\footnote{Relatively weak [OII] emission 
  is predicted for a hard radiation field \citep{ho2005}.}, 
we can additionally estimate 
\begin{equation}
SFR_4(M_\odot \textnormal{  yr}^{-1})=(1.4\pm 0.4)\times 10^{-41}L([\mbox{\ion{O}{2}}])(\textnormal{ergs s}^{-1})^{-1}=1.25 \pm 0.36 \ M_\odot\textnormal{  yr}^{-1} . 
\end{equation} 
The two estimates are in good agreement, suggesting that dust is not
adversely affecting the line fluxes. 
Of course, we are not sensitive to any highly obscured star formation
and in this sense the measured star formation rate is a lower
limit. We also note that this SFR is only measured over bin 4g. We
also detect H$\alpha$ emission for the galaxy within bin 3g but we 
expect that these line fluxes are influenced by the neighboring
QSO's ionizing radiation. Therefore, we assume that the star formation in
that bin is equal to bin 4g and apply a multiplicative
correction factor of 2. We also need to correct for the fact that our slit is only 1$''$ wide
and the galaxy extends outside of the slit. To do this, we make use of
our $R$-band image and (1) measure the total flux of the galaxy (2) apply a 1$''$ slitmask
and measure the total flux which falls within this slit. The ratio of these two fluxes is 
the slit loss correction. We find this to be a factor of 2.5. We thus apply a total
correction factor of 5. This of course is assuming that the star formation
rate is constant across the entire galaxy and that the $R$-band light follows the star formation.
 This gives an estimate of the total star
formation rate for the galaxy of $8.05\pm 0.15 M_\odot \mbox{
  yr}^{-1}$. 

As a caveat, we point out that the star formation in the
companion galaxy may be dominated by its innermost regions and thus
not indicative of the overall galaxy morphology. This may occur when
gas is funneled by tidal torques of
the interaction to trigger a burst of recent star formation. If this is the case, 
we may be overestimating the slit loss correction.  For a tabulated summary of
all SFR estimates with the varying assumptions, see Table \ref{table:all}. These estimates
likely bracket the range of possible star formation rates.

We may also estimate the metallicity of the gas in bin~4g using
standard line flux ratio calibrations. \cite{pp04} outline two
calibrations of metallicity determination. In these units solar O/H is 8.66 \citep{solarO1, solarO2}: 
\begin{eqnarray}
12+\log(\mbox{O/H})&=&8.9+0.57\times N2 = 8.64 \pm 0.01 \\
12+\log(\mbox{O/H})&=&8.73-0.32\times O3N2 = 8.49 \pm 0.01
\end{eqnarray}
where $N2\equiv\log \{[$\ion{N}{2}$]/\mbox{H}\alpha\}$ and $O3N2\equiv\log \{\left([\mbox{\ion{O}{3}}]/\mbox{H}\beta\right)/ \left( [\mbox{\ion{N}{2}}]/\mbox{H}\alpha\right)\}$.
These empirical relations have scatter such that 95\% of points are within 0.38 dex of the $N2$ calibration and within 0.25 dex of the $O3N2$ calibration.  Thus the two
measurements are in agreement within the systematics of each other and the solar value.
Therefore, we conclude that the galaxy has roughly solar metallicity.

\subsubsection{Mass Estimates}\label{sec:galmass}
We can estimate the mass of the companion galaxy in the same manner as for the quasar host
galaxy ($\S$\ref{sec:bhprop}), i.e., through analysis of the rotation curve. 
The measured circular velocity implies an estimate to the dark matter halo mass 
 $M_{DM}>2.1\times 10^{11} M_\odot$.
This is also a lower limit due to uncertainties in the inclination,
axis ratio, and position angle of the galaxy. We note, however, that
unlike the quasar host, the data does extend to the flat portion of
the rotation curve. 

\label{sec:galstellar}
Using the 5 SDSS colors ($ugriz$), one may constrain the spectral
energy distribution of a galaxy at $z<1$ to 
estimate its mass-to-light ratio. This mass-to-light ratio can then be combined with the observed luminosity to estimate its stellar mass.
Making use of version v4.1.4 of the \textsc{kcorrect} code  \citep{kcorrect}, we estimate a stellar mass of $1.9\times10^{10}M_\odot$ with a typical error
of approximately 0.2 dex compared to other photometric estimates of
stellar mass. Using the calibration of \cite{conroy2009} which is very similar to that of \cite{fontana2006}, 
we find that $M^{*}(z=0.369)=1.01 \times 10^{11} M_\odot$. Therefore,
the companion galaxy appears to be $\approx 0.2 M^{*}$.
Using halo abundance matching to relate stellar mass to halo mass
\citep[e.g.][]{conroy2009}, we estimate the 
dark matter halo mass to be approximately $6\times 10^{11} M_{\odot}.$

\begin{figure}
\epsscale{1.5}
\plotone{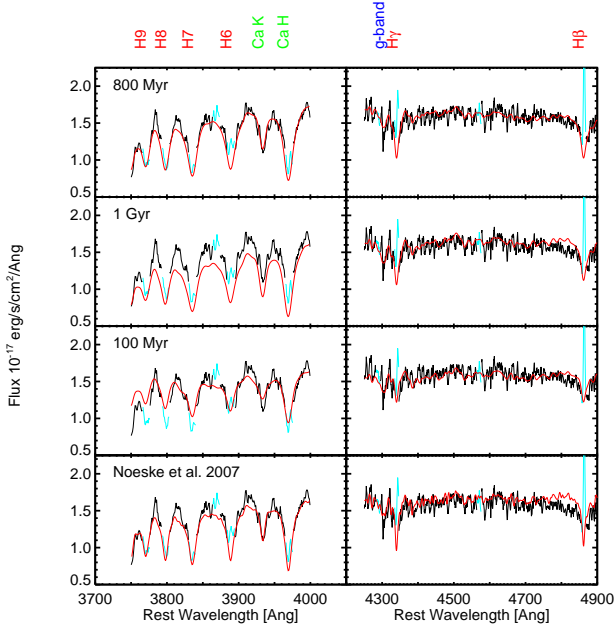}
\caption{ Comparison of SSP models to the observed spectrum. Data are
  shown in black, the models are shown in red, and masked emission
  line regions are shown in cyan. The observed strong Balmer
  absorption lines are characteristic of A star features which
  indicate star formation halted in a time period that is longer than
  the lifetime of O and B stars but shorter than that of A stars. This
  range is 100 Myr - 1 Gyr and we refined the age estimate with SSP
  modeling. We find that the range of reasonable models is for
  starbursts that started t=300-800 Myr ago. Older stellar populations
  do not have strong enough Balmer features and are too red. Younger
  populations also underestimate the Balmer features and are too
  blue. Our best model is plotted in the first row which corresponds
  to 800 Myr. The second and third rows show examples of the poor fits
  that result from too old and too young populations respectively. We also show
  in the bottom panel the results of using a star formation history determined
  by \cite{noeske07}. This SFH insufficiently reproduces the spectral region between 
  Ca K and H7.
}
\label{fig:ssp}
\end{figure}

\subsubsection{Stellar Population Models: Is the Companion a Post-Starburst Galaxy? }\label{sec:stellarpops}
The classic definition of a post-starburst galaxy is one which has strong Balmer absorption features along with
Ca H+K absorption lines and an absence of emission lines.
While our galaxy does have emission lines which point to ongoing
star formation, we believe that our galaxy falls into the post-starburst class
for two reasons: (1) a significant portion of the emission may be a result from photoionization
from the nearby quasar and (2) our following analysis demonstrates that 
the star formation was likely elevated in the past in order to 
simultaneously reproduce the colors and absorption lines.

We utilize \cite{bc03} population synthesis models to constrain the star formation history of this galaxy.
We construct solar metallicity models using Padova 1994 stellar evolutionary tracks with a \cite{chab} IMF. 
Following \cite{yan2006},
we used various two-component  models for the burst star formation history. The first component is 
a single 7 Gyr old, instantaneous stellar population which represents
an old passively evolving background stellar population.  The second
component is a $\tau$ model burst occurring within the last 2 Gyr.
We used a variety of $\tau$'s to parameterize our burst of star
formation [where $SFR\propto \exp(-t/\tau)$]:
0.01, 0.1, 0.25, 0.5, 0.75, 0.9, 2, and 4 Gyr and spaced the age of the burst at 100 Myr intervals. 
We allowed a range of burst amplitudes ($A$; defined in the same manner as \cite{kauffmann2003}) such that the stellar mass  produced by the burst 
(evaluated at $t=\infty$) relative to the initial 7 Gyr population was between $10^{-3}$ and $10^4$ in logarithmic steps of $\Delta \log A = 0.53 $. 

We fit these model spectra to match our observed spectrum after correcting for Milky 
Way dust absorption in the same manner we treated our emission flux measurements (see $\S$ \ref{sec:linefluxes}).
Due to the LRIS dichroic gap
we restricted ourselves to two rest wavelength ranges $3700-4000$ \AA \ and $4250-4900$ \AA. 
We also masked all pixels within 400 km/s of an emission line since our models do not include nebular emission. 
We allowed the model spectrum to scale up or down
by a single constant factor and allowed for an intrinsic dust correction. The dust correction was implemented 
with a \cite{charlot_fall2000} law parameterized by $\tau_V$:
\begin{equation}
 F_{\lambda,obs} = F_{\lambda,int} \ \textrm{ exp}\left[-\tau_V  \left( \frac{\lambda }{ \mbox{5500 \AA}}\right)^{-0.7}\right]
\end{equation}
where $F_{\lambda,obs}$ is the observed flux and $F_{\lambda,int}$ is the intrinsic flux.

We applied the additional criterion that the SDSS $u-g$, $g-r$, $r-i$, and $i-z$ colors of the models (including
applying the Milky Way dust absorption and fitted intrinsic dust) are within 0.5 mag of the observed colors.

Altogether, we created 2016 models. We find that no individual model
provides a good ($\chi^2_r=1$) fit to all of the spectral features. 
However the models that simultaneously best reproduce the observed colors and the line features (both the Balmer and Ca H+K)
 have ages for the burst component that fall in the range of 300 - 800 Myr,  star formation histories
 parameterized by  0.25 Gyr $<\tau<$ 0.9 Gyr, and have a burst
 mass fraction (relative to the old passively evolving component) that fall in the range  0.009-0.1 (see Fig.~\ref{fig:ssp}). 
 However we note that among those models there are a handful which have a mass fraction as high as 4.
 All models have dust reddening characterized by
 $\tau_V<1.01$ with very few having $\tau_V >0.7$. The best model has $\tau_V=0$,
consistent with the observed H$\alpha$/H$\beta$ ratio
(Figure~\ref{fig:lineratios}).
 Older models are too red, while younger models are too blue.
 Higher burst fractions underestimate Ca H+K and g-band absorption features while lower burst fractions have
 insufficient Balmer features or too red colors. In summary, we find the data are best described by a substantial
  burst that occurred 800 Myr ago and that models
 with $t<300$ Myr and $t>800$ Myr are disfavored.

A reasonable question to ask is whether the companion galaxy is consistent with a
normal, non-burst galaxy. To do this we make use of the star formation
histories derived by \cite{noeske07} for star-forming galaxies at the redshift of our system. They present
parameters of $\tau$ model star formation histories as a function of baryonic mass.
We have an estimate of the stellar mass of the galaxy (see above) which we assume to be
the baryonic mass of the galaxy; corrections to include gas up to reasonable gas fractions ($\lesssim50\%$)
have little to no effect on the following results. Thus, we find
the star formation parameters to be $\tau=25.06$ Gyr and after converting from redshift intervals
to ages, we find the beginning of the model be $5.13$ Gyr prior to our current
observation of the galaxy. Thus this model is very nearly
a constant star formation rate model. 
We fit this model to the data allowing for a normalization factor and dust reddening
as we did for the burst models. We present the results in the bottom panel of Fig.~\ref{fig:ssp}. One
can see that this model insufficiently produces the spectral features between Ca K and H7. 
Additionally, the model's large amount of ongoing star formation produces
too blue of an SED and thus requires a rather large column of dust to 
redden the spectrum ($\tau_V=1.27$). Assuming the same amount of reddening corresponds
to the \ion{H}{2} regions that we use to estimate the SFR, we find two results. Firstly, this
measure of reddening implies a higher H$\alpha$/H$\beta$ ratio than
observed, but only by $\gtrsim 1\sigma$. Secondly, the reddening
correction would increase our estimate of the SFR by
a factor of 3.7 giving a value greater than 20 $M_\odot/$yr. 
Comparison with the observational results of
the specific star formation sequence \citep{noeske07_2}, we find
this star formation rate lies well above the 84th percentile for galaxies at this stellar mass.
These pieces of evidence 
argue against this scenario, but
we cannot rule it out altogether.
 
 The relatively high level of ongoing star formation may also be an indicator
 of a previous phase of elevated star formation. SFHs for galaxy mergers 
 are not $\delta$-functions. For example, 
 \cite{cox2006} show that while the SF activity reaches a peak 
 after first passage it can remain elevated until the merger
 ultimately coalesces. 
 
\section{Analysis of the Bridge}
\subsection{Density \& Mass }\label{sec:column}

We can make an order of magnitude estimate of the column
 density along the line of sight through the bridge by using the
observed luminosity of the bridge in recombination lines
 combined with a simple volume model for the bridge (See Fig.~\ref{fig:cartoon}). 
We note that the column density we are about to estimate 
is \emph{not} the column density of material that the quasar radiation encounters
 along the bridge, but is the projected column on the sky from our perspective.
  We denote this column density as $N_{H,\perp}$ because if the bridge is in the plane of
the sky then this column density is perpendicular to the column density that the quasar's radiation encounters.

We start with the expression for the luminosity of the H$\alpha$ line produced by recombinations:
\begin{equation}\label{eqt:lha}
L_{\textrm{H}\alpha}=n_e n_p \alpha_{\textrm{H}\alpha}^{eff} Vfh\nu_{\textrm{H}\alpha}
\end{equation}
where $L_{\textrm{H}\alpha}$ is the luminosity of H$\alpha$ in a given spatial region, $n_e$ is the number 
density of electrons, $n_p$ is the number density of protons, $\alpha_{\textrm{H}\alpha}^{eff}$ is the $\textrm{H}\alpha$ effective 
recombination coefficient, $V$ is the volume the gas fills,  $f$ is the filling factor of the gas in that volume (observationally 
unconstrained but by definition $\le 1$), $h$ is 
Planck's constant, and $\nu_{H\alpha}$ is the frequency of the H$\alpha$ transition.
We assume that the bridge is a cylinder with diameter $\mathcal{D}$ that extends a
radial distance $r_1$ from the QSO  to a larger radial distance $r_2$ (we note that $r_1,r_2 \gg \mathcal{D}$;
see Fig.~\ref{fig:cartoon}).  In this case, $V\approx\mathcal{D}^2\left(r_2 -r_1\right)$.

Assuming that the hydrogen is fully ionized, we take $n_e=n_p=n_{H}$, 
where $n_H$ is the number density of hydrogen in all ionization states. This
is a reasonable assumption given the observed line
ratios\footnote{Another piece of evidence that the gas is mostly
  ionized is from the observation that the emission extends along a
  long spatial extent. If the quasar is responsible for the ionization
  ($\S$\ref{sec:source}) the gas must therefore be optically thin to
  ionizing radiation along a pathlength of order $r_2$. The optical
  depth to ionizing photons for {\bf neutral} hydrogen reaches unity for
  column densities of $\sim 10^{17}\mbox{ cm}^{-2}$. Since our column
  density is measured along a pathlength of $\mathcal{D} \ll r_2$, as
  long as we measure a total hydrogen column density (\ion{H}{1} and \ion{H}{2}) of  $\gtrsim 10^{17}\mbox{
    cm}^{-2}$ then this is a safe assumption. If there is more than enough gas to be optically thick
    if not ionized and the gas is {\bf not} optically thick, then the gas must be ionized.}.

Manipulating eqn.~\ref{eqt:lha} and substituting for our expression of $V$ we define:
\begin{equation}
\omega \equiv f^{1/2}n_H\mathcal{D}=\sqrt{\frac{L_{H\alpha}}{\left(r_2-r_1\right)\alpha_{H\alpha}^{eff}h\nu_{H\alpha}}}
	\approx     10^{21}  \mbox{ cm}^{-2}.
\end{equation}
This quantity consists only of observed quantities and atomic constants. The order of magnitude estimate is made using the 
observed flux of H$\alpha$ for bin 1b which we round to $4\times10^{-17}$ erg/s/cm$^2$. Using the observations
from bin 1b, we put a limit on the volume density of the gas by noting that:
\begin{equation}
n_H \approx  \frac{\omega}{ f^{1/2} \mathcal{D}} \approx 1  \  f^{-1/2} \left(\frac{\mathcal{D}}{1 \mbox{ kpc}}\right)^{-1} \mbox{ cm}^{-3}
\end{equation}
Taking the limit to $\mathcal{D}$ placed by our Kast observations ($<$ 10 kpc), we derive 
$n_H \gtrapprox 0.1 \ f^{-1/2} \mbox{ cm}^{-3}$. This is consistent
with volume densities estimated for similar gaseous bridge structures in 
galaxy merger simulations by \cite{weniger}, where they
implemented a multiphase ISM code in a galaxy merger simulation.

We may now estimate the column density,
\begin{eqnarray}
N_{H,\perp}&=&\frac{\mbox{ $\#$ of Hydrogen atoms}}{\mbox{Surface Area those atoms fill}}\approx 
	\frac{fn_HV}{\mathcal{D}\left(r_2-r_1\right)}\nonumber\\
	&\approx& n_H f \mathcal{D}\approx \omega f^{1/2}  \approx f^{1/2} \ 10^{21}  \mbox{ cm}^{-2},
\end{eqnarray}
This value is consistent with that of 21 cm neutral hydrogen
observations of tidal tails of the Antennae galaxy by \cite{hib01}. If
that gas were fully ionized, then it would give rise to a similar
column density of H$^+$. 
Lastly, we estimate the hydrogen mass
\begin{eqnarray}
M_{H}&=& n_H V f m_p \approx  \mathcal{D}\left(r_2-r_1\right) \omega f^{1/2} m_p \nonumber\\
  &\approx&   	10^8 \ f^{1/2} \left(\frac{\mathcal{D}}{1 \mbox{ kpc}}\right) M_\odot.
\end{eqnarray}
This is relatively modest, but we note that $\mathcal{D}$ could be as large as
10\,kpc although $f$ could also be significantly smaller than unity.

\begin{figure}
\epsscale{1.2}
\plotone{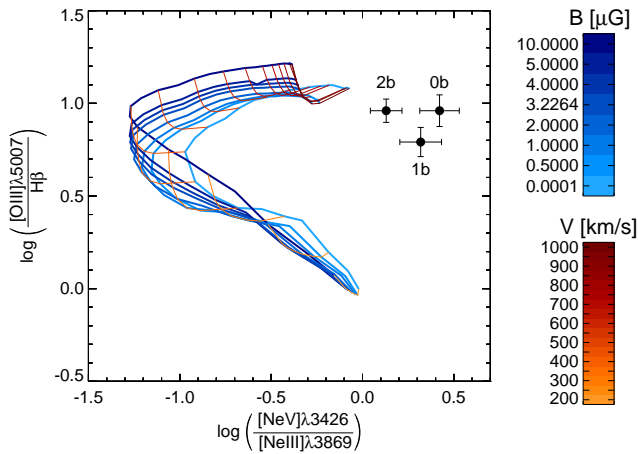}
\caption{Comparison of the predicted line ratios for fast, radiative
  shocks \citep[curves;][]{shock_models} with the observed values in
  the bridge (bins 0b, 1b, and 2b).  These predictions are for fast
  radiative shocks+precursor in gas with a density of 1 cm$^{-2}$ and
  solar metallicity. The range spanned by the predictions is
  relatively insensitive to these choices.  The bridge's emission is
  inconsistent with shocks even if we allow those shocks to be
  unreasonably fast (up to 1000 km/s).
  }
\label{fig:shocks}
\end{figure}

\subsection{Source of Ionization}\label{sec:source}
We observe the bridge in emission lines from recombination of Hydrogen and forbidden
lines of high ionization species. This, of course, requires a physical source of ionization and heating of the gas.
In this section, we compare our observed line ratios against models to determine the 
source of ionization. After ruling out collisional ionization and photoionization from fast radiative shocks and \ion{H}{2}
regions, we conclude that the quasar is shining on the bridge and photoionizing the gas and test this hypothesis using \textsc{cloudy} 
models.

\subsubsection{Shock Ionization}

Shocks are common occurrences in galaxy mergers and also result from the interaction 
of quasar jets with the interstellar and circumgalactic media of the galaxy 
\citep[e.g.][]{rosario10}. If those shocks are fast enough, they could ionize
gas along the bridge.
There are two ways in which shocks can lead to ionization:
(1) collisional ionization of the shocked gas or (2) radiative shocks that 
photoionize  the pre-shock gas with the radiation emitted by the post-shock gas. 
We begin by addressing the
first possibility: whether gas in the bridge has been collisionally ionized.

If the gas were collisionally ionized, then emission from high ionization potential states such as \ion{Ne}{5}, 
 would require a very large electron temperature. 
We can estimate $T_e$ using the [\ion{O}{3}] transitions as these form a temperature diagnostic in the density regime $n_e \le 10^5 \mbox{ cm}^{-3}$  \citep{agn2}:
\begin{equation}
\frac{j_{\lambda 4959}+j_{\lambda 5007}}{j_{\lambda 4363}}=\frac{7.9\,\textrm{exp}(3.29\times10^4/T_e)}{1+4.5\times 10^{-4}n_e/T_e^{1/2}} .
\end{equation}
This diagnostic is valid as long as the density is below the critical density of the [\ion{O}{3}] transitions. If the density were higher than this critical density, the flux would be attenuated dramatically since at these densities collisional deexcitation dominates over radiative deexcitation. Such high density, however, would imply much lower
[\ion{O}{3}]/H$\beta$ ratios than observed. Such a density is also many orders of magnitude higher than normal ISM densities and is highly unlikely.
The conservative upper limit to the  [\ion{O}{3}]$ \lambda 4363$ line flux
sets an upper limit to the gas temperature of
$T_e<7\times 10^4$ K. This is over 4 times lower
than the characteristic temperature 
of \ion{Ne}{5} \citep[$3\times10^5$ K;][]{neV}. 
The detection of [\ion{Ne}{5}]$\lambda \lambda$3426,3346 combined
with the temperature upper limit from [\ion{O}{3}]
rules out the hypothesis that  the primary ionization source is collisional ionization.

Radiative shocks, via photoionization, can produce 
high ionization lines in conditions with lower electron temperatures. 
In such shocks, the post-shock gas reaches high temperatures and its emission photoionizes 
the pre-shock gas. This produces a gas that is ionized by a relatively 
hard spectrum. However, the observed emission line ratios 
in the bridge are inconsistent 
with those given by the diagnostic diagrams of \cite{shock_models}
 for fast radiative shock+precursor models that span shock velocities 
 of 100-1000 km/s, magnetic paramters $B/\sqrt{n}=10^{-4} - 10 \, \mu 
  \mbox{G cm}^{3/2}$, $n=0.01-1000 \mbox{ cm}^{-3}$, and a variety of 
 abundance ratios. These models take into
 account both emission of the shocked material and emission from 
the pre-shock gas.
The line ratios that are clearly discrepant with our observations include the
[\ion{O}{2}]$\lambda 3727/$[\ion{O}{3}]$\lambda 5007$ and
  [\ion{Ne}{5}] $\lambda 3426/$[\ion{Ne}{3}] $\lambda 3869$ ratios.
Most striking is that all of these models predict
 [\ion{Ne}{5}]$\lambda 3426$ fainter than [\ion{Ne}{3}] $\lambda 3869$,
in direct contradiction
 with our observations of the bridge (see Fig.~\ref{fig:shocks}).

Lastly, we stress that the bridge emission extends along the entire length of the $>38 h_{72}^{-1}$
 kpc bridge connecting the two galaxies. It could be difficult to simultaneously generate shocks 
across such a large spatial extent. 
Furthermore, if we assume that the beginning of this interaction (and hence when the shocking began) is 
 consistent with the age that we estimate from the stellar population
 modelling, this could require shocks to be sustained 
over a very long spatial extent and for a very long time ($300-800$ Myr).

Using all of the above reasoning, we rule out shocks as a viable mechanism
for the ionization of the bridge.
 
\subsubsection{\ion{H}{2} Regions}\label{sec:h2}
Studies of merging systems often reveal evidence for new star formation,
and therefore elevated populations of young and massive stars.
These star-forming regions are expected to be surrounded by \ion{H}{2}
regions of ionized gas which could produce the line emission observed
in the bridge.  We have tested this hypothesis in the following manner.

Looking at the BPT diagram (Fig.~\ref{fig:bpt}) as first presented in
$\S$~\ref{sec:metal}, we 
see that the spatial bins that contain gas in the
bridge (0b, 1b, 2b), show the bridge emission line ratios are
consistent with gas ionized by an AGN spectrum.
The gas in these bins also emits [\ion{Ne}{5}] lines;
 common tracers of AGN photoionization \citep{nev_agn}.
In fact the detection of  \ion{Ne}{5} strongly rules out \ion{H}{2} regions as
the dominant ionization source because
stars lack a hard enough spectrum to effectively ionize Ne to this high of an
ionization state. 

Looking at the two bins which cover the galaxy (3g and 4g), 
the emission characteristics vary dramatically across 
the two regions. While bin 4g (the far-side of the galaxy)
 has a strong contribution 
from \ion{H}{2} regions, bin 3g (the near side of the galaxy) 
falls somewhat in between the regions of AGN and \ion{H}{2} region ionization.
This abrupt change in emission characteristics 
is consistent with a model where the quasar is 
photoionizing the gas in the bridge and the near side of the galaxy 
but where the galaxy
effectively shields its own far 
side from the radiation. 
The abrupt change of the emission characteristics of bin 3g from those of the bridge is likely the result
of the gas changing from the optically thin conditions of bins 0b, 1b, and 2b to the
optically thick conditions of the disk of the galaxy in bin 3g. Bin 3g, also likely has some
contribution from \ion{H}{2} regions, complicating its analysis.

\begin{figure}
\epsscale{1}
\plotone{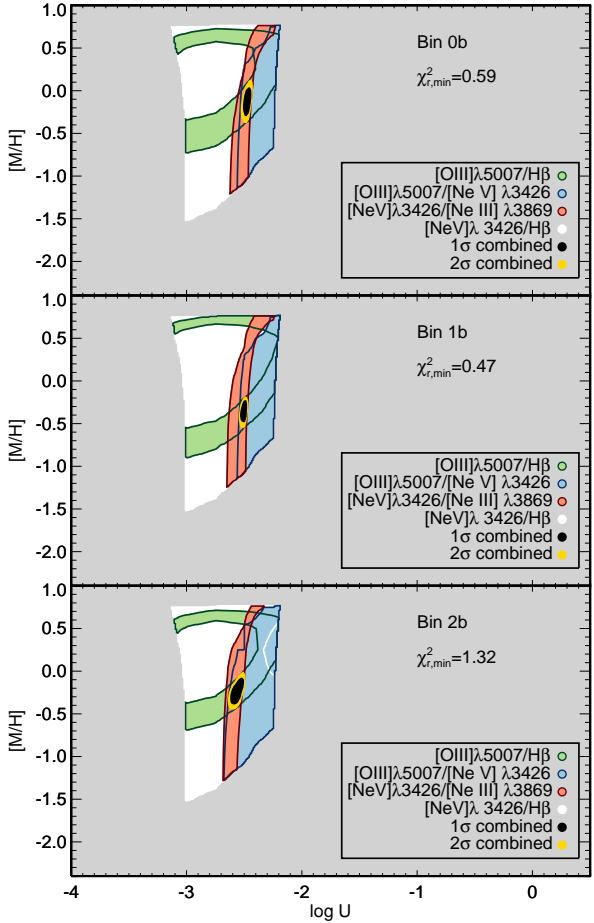}
\caption{Plots revealing the allowed parameter space in ionization
  parameter $U$ and metallicity [M/H] for the emission observed in the
  bridge (bins 1b and 2b). The grey region in each plot is excluded
  because these models predict significant emission from lines that
  are not detected at that level (e.g., \ion{Fe}{7}, \ion{O}{6}, \ion{O}{2}). The
  colored areas in each figure represent the $1\sigma$ $\chi^2$
  contours for a series of line ratios as labelled. The black and
  yellow contours correspond respectively to the 1$\sigma$ and
  2$\sigma$ confidence intervals for the combined $\chi^2$ contours of
  all 4 line ratios. Note that the entire region which is not grey is
  allowed at $1\sigma$ confidence for the [\ion{Ne}{5}]$\lambda
  3462$/H$\beta$ ratio for bins 0b and 1b. Table~\ref{table:all} lists
  the best fit values.
} 
\label{fig:chi2}
\end{figure}

\subsubsection{Quasar Ionization Modelling \& Bridge Metallicity}\label{sec:cloudy}

In the previous sections we ruled out shocks and stellar photoionization as the source
of ionization for the bridge material. We also inferred, from investigation of the BPT
diagram, that the line ratios were likely the result of photoionization by a quasar. In this section, we test this hypothesis
by comparing our observations with photoionization models.

Emission line ratios can inform one about the physical conditions of a photoionized gas as well as the
properties of the incident ionizing radiation. 
We utilize \textsc{cloudy}  version 08.00 photoionization simulations \citep{cloudy} to
determine if photoionization by a quasar's radiation field can
reproduce our observed line ratios and then we infer physical conditions
from these ratios.

Our observations include a variety of line 
ratios for this purpose.  We focus on lines with relatively high
signal to noise and ratios that 
depend sensitively on ionization state and metallicity:
[\ion{O}{3}]$\lambda$5007/H$\beta$,
[\ion{O}{3}]$\lambda$5007/[\ion{Ne}{5}]$\lambda$3426,
[\ion{Ne}{5}]$\lambda$3426/H$\beta$, and
[\ion{Ne}{5}]$\lambda$3426/[\ion{Ne}{3}]$\lambda$3869.  
By studying line ratios instead of absolute luminosities, 
the results are insenstiive to the detailed geometry
of the system.

Our models are plane-parallel slabs of gas illuminated by an ionizing spectrum. 
We parameterize the shape of the ionizing continuum as a power law with a spectral index of $\alpha=-0.4$ from 0.01 to 20 Rydberg (consistent with
results from modelling the QSO SED as described in Paper II).
 The results are not very sensitive to changes
in the spectral range the power law covers. 
 We included the appropriate CMB 
 as well as cosmic rays, but 
these sources of ionization have little consequence for our calculations.
We ran models with metallicities of $-2.5<$[Fe/H]$<1.0$ . We note that our implementation of metallicity is just a scaling of all elements equally relative to solar
and hence [Fe/H] can
be interpreted as a generic [M/H], [O/H], or any [X/H].
We varied the strength of the ionizing radiation by varying the ionization parameter $U$ which is defined as the ratio of
the ionizing photon density to hydrogen density:
\begin{equation}
\log U=   \log \left(\frac{S}{ 4\pi r^2 cn}\right) 
\end{equation} 
\noindent where $r$ is the distance from the ionizing source, $n$ is the number density of the gas, and $S$ is the ionizing photon luminosity of the source. 
We set the hydrogen density to be $10 \, \mbox{cm}^{-3}$ (however the particular value was not
relevant because we varied the ionization parameter $U$, and optically
thin gas is homologous in $U$). 
We chose the gas stopping column density to be such that the gas was optically thin ($N_{HI}=16.8$).
We have adopted the optically thin regime 
because we see ionization extending throughout the entire 38~\hkpc\
extent of the bridge. 
Optically thick models would need to
be clumpy and only partially covering the source. 
We discuss futher the implications of clumpy models in Paper~II.

\begin{deluxetable*}{lll} 
\tablecaption{Measured \& Derived Properties}
\tablehead{\colhead{Property} & \colhead{Estimated Value} &
  \colhead{Section}}
\startdata
\cutinhead{Merger}
$z_{QSO}$ & $0.369\pm0.001$ & \ref{sec:1dspectra}\\
observed $\Delta v$   & $30\pm 30$ km/s   & \ref{sec:kinematics}\\
Projected Distance & 38 $h_{72}^{-1}$ kpc & \ref{sec:2dspectra}\\
Merger Stage & Between First and Second Passage & \ref{sec:discussion}\\
\hline
\cutinhead{Quasar}
$M_{BH}$  & $10^{8.7}M_\odot$   &\ref{sec:bhprop}; \cite{shen08}\\
$L_{Bol}$  &  $10^{45.7}$ erg/s    &\ref{sec:bhprop}; \cite{shen08}\\
Eddington Ratio & 0.07 & \ref{sec:bhprop}; \cite{shen08}\\
Host DM Mass & $>8.8\times 10^{11} M_{\odot}$ & \ref{sec:bhprop}\\
\cutinhead{Companion Galaxy}
Concentration & $>$2.61 & \ref{sec:imaging}\\
SFR$_4$ & $1.61\pm 0.03 M_\odot$ yr$^{-1}$ & \ref{sec:metal}\\
Assuming SFR$_3$=SFR$_4; SFR$ & $3.22\pm 0.06 M_\odot$ yr$^{-1}$ & \ref{sec:metal}\\
Assuming R-band Light traces SF; SFR & $8.05 \pm 0.15 M_\odot$ yr$^{-1}$ & \ref{sec:metal}\\
$12+\log (\mbox{O}/\mbox{H})$ & $8.64 \pm 0.01 \pm 0.2$ & \ref{sec:metal}\\
$M_{DM}$ & $>2.1\times 10^{11} M_\odot$ & \ref{sec:galmass}\\
$M_\star$ & $1.9\times 10^{10} M_\odot$ & \ref{sec:galmass}\\
SHAM $M_{DM}$ & $6\times10^{11} M_\odot$ & \ref{sec:galmass}\\
SSP Age & 300-800 Myr & \ref{sec:stellarpops}\\
Burst Mass Fraction & 1-100\% & \ref{sec:stellarpops}\\
\cutinhead{Bridge}
Kinematics & See Fig.~\ref{fig:kinematics} &\ref{sec:kinematics}\\
$\mathcal{D}$ & $<10.2 h_{72}^{-1}$ kpc & \ref{sec:2dspectra}\\
$n_H$ & $\approx  f^{-1/2} \left(\frac{\mathcal{D}}{1 \rm kpc}\right)^{-1} \mbox{ cm}^{-3}$ &\ref{sec:column}\\
$N_{H,\perp}$ & $\approx f^{1/2} \ 10^{21}  \mbox{ cm}^{-2}$ & \ref{sec:column}\\
$M_H$ & $\approx  10^8 \ f^{1/2} \left(\frac{\mathcal{D}}{1 \rm kpc}\right)  M_\odot $ & \ref{sec:column}\\
Source of Ionization & QSO Photoionization & \ref{sec:source}\\
Bin 0b $\log U$ &  $-2.48_{-0.02}^{+0.03}$ & \ref{sec:cloudy}\\
Bin 0b [M/H]&  $-0.13_{-0.16}^{+0.16}$ & \ref{sec:cloudy}\\
Bin 0b $\chi^2_r$ & 0.59 &\ref{sec:cloudy}\\
Bin 1b $\log U$ &  $-2.50_{-0.02}^{+0.03}$ & \ref{sec:cloudy}\\
Bin 1b [M/H]&  $-0.36_{-0.12}^{+0.12}$ & \ref{sec:cloudy}\\
Bin 1b $\chi^2_r$ & 0.47 &\ref{sec:cloudy}\\
Bin 2b $\log U$ &  $-2.56^{+0.05}_{-0.05}$  & \ref{sec:cloudy}\\
Bin 2b [M/H]  & $-0.23^{+0.14}_{-0.16}$  & \ref{sec:cloudy}\\
Bin 2b $\chi^2_r$ & 1.32 &\ref{sec:cloudy}\\
\enddata
\label{table:all}
\end{deluxetable*}

We compared our line ratios to the models using a $\chi^2$ statistic which requires
careful estimation of the errors on those line ratios. We refined our error estimates by including an additional 10 percent error 
on lines if they are part of a line ratio that spanned different cameras to account
for relative fluxing issues. The models do not take into account any dust that may be a 
result of an intervening screen of material between the region of emission and the observer.
To characterize the error associated with this, we turned to our measured line ratios.
Specifically, we characterized our errors
from reddening through 1000 Monte Carlo realizations of the observed 
H$\alpha$/H$\beta$ ratio. We adopted the observed ratio
with a normally distributed error to determine the intrinsic
reddening and then corrected the line ratios. This created 1000 ``actual"
 line ratios whose standard deviation we used to estimate the error due to 
intrinsic reddening. The log relative errors on  ([\ion{O}{3}]$\lambda$5007/H$\beta$, 
[\ion{Ne}{5}]$\lambda$3426/[\ion{Ne}{3}]$\lambda$3869,
[\ion{Ne}{5}]$\lambda$3426/H$\beta$,
[\ion{O}{3}]$\lambda$5007/[\ion{Ne}{5}]$\lambda$3426) were (-1.4,-0.9,-0.12,-0.62) for 0b,
(-1.8,-1.5,-0.8,-0.7) for bin 1b, and (-4.1,-3.7,-3.0,-3.05) for bin~2b. The relative errors are much
lower for bin 2b, because the observation places it slightly below the theoretical value.

We first ruled out any models that predicted detectable fluxes (30\%
of the luminosity of [\ion{O}{3}]$\lambda 5007$) for high ionization
lines (e.g. \ion{Fe}{7} 6087, \ion{O}{6} 5291, etc.) or low ionization lines
(e.g. \ion{O}{2}) which are not detected at that level in the bridge. 
This rules out low and high ionization paramters ($\log U<-3; \log U >
-2$) where [OIII]~5007 is predicted to be weak relative to these lines.
%
This resulted in the excluded region denoted grey in Fig.~\ref{fig:chi2}. 
With well-characterized errors and the suite of models in hand, we
next constructed $\chi^2$ contour surfaces as a function of both $U$ and [M/H]. 
Here we see that the 
[\ion{Ne}{5}]$\lambda$3426/[\ion{Ne}{3}]$\lambda$3869 constrains $\log U$ to a narrow region.
We see that our best constraint on [M/H] comes from [\ion{O}{3}]$\lambda$5007/H$\beta$. The 
intersection of these two allow a that region is double valued.  The overall $\chi^2$ and the [\ion{O}{3}]$\lambda$5007/[\ion{Ne}{5}]
ratio, however, break this degeneracy. Due to lower signal-to-noise, the [\ion{Ne}{5}]$\lambda$3426/H$\beta$
ratio offers little additional constraint. The combined constraint lies in the same region for
all three bins and allows us to 
estimate both [M/H] and $\log U$ with relatively high precision
(summarized in Table~\ref{table:all}). 

The best fit models for bin 0b, bin 1b and bin 2b had $\chi^2_r$ of 0.59, 0.47 and 1.32 respectively and
predict [\ion{O}{3}]$\lambda$4363/ [\ion{O}{3}]$\lambda$5007=0.03
which is well below our detection limit. Therefore these models 
reproduce the observed [\ion{Ne}{5}] emission at a low
electron temperature ($T_e \sim 1.8 \times 10^4$\,K). 
There are additional systematic errors associated with the simple
geometry in \textsc{cloudy}, the presumption of equilibrium, and 
uncertainty in the ionization spectral shape (e.g.\ variations in the
spectral index of the power law $\alpha$).  We characterize a
systematic error based on variations of the power law index of a few tenths
to produce an
additional uncertainty on $\log U$ of $\approx 0.05$\,dex and on [M/H]
of $\approx 0.06$\,dex. 
%

Lastly, we need to determine if the derived 
values for the material make physical sense.
 Given that a quasar can output $10^{56}$ ionizing photons per second \citep{tad96} and the
 bridge is approximately 20 kpc away, this yields a possible $U$ parameter of $\log U=-1.16-\log n$. Noting that
 the QSO could always be partially obscured or non-isotropic, the quasar is a feasible source
  for any density greater than
 $\log n \approx -1.5 \mbox{ cm}^{-3}$. We also point out that our
 estimates for $U$ 
 do not appear to fall off as $1/r^2$ from the quasar. In Paper II, we discuss
 various possible interpretations of this observation.

\begin{deluxetable*}{lllc} 
\tablecaption{Comparison with Theory}
\tabletypesize{\footnotesize}
\tablehead{\colhead{Theoretical Prediction} &\colhead{Reference}& \colhead{Observation} &  \colhead{Consistent?}  }
\startdata
Galaxy internal parameters &\cite{mihos96} & One galaxy is a quasar, &Y \\
determine their evolution in a merger& &one has no AGN activity &\\
\\[.1 mm]
 First passage inspires a starburst     &  \cite{mihos96} & Starburst spectrum observed & Y \\
 & & after first passage, before second& \\
\\[.1 mm]
 Mergers may lead to quasars & \cite{hop08}     &  One galaxy observed as a quasar  & Y \\
 \\[.1 mm]
 This quasar phase typically & \cite{hop08} & Merger between first&?\\
 occurs during final coalescence & & and second passage&\\
 \\[.1 mm]
  Starburst \& tidal effects may & \cite{stockton1982} & Imaging is mildly concentrated,     & ? \\
lead to a concentrated morphology &&   but inconclusive due to  & \\
&&insufficient spatial resolution& \\
\\[.1 mm]
  2 Gyr timescale between first  &\cite{lotz_mass}&  $t=300-800$ Myr after first& Y \\
  and second passage && passage, before second passage&\\
  \\[.1 mm]
  Kinematics near apogee have     &\cite{galmer} & $\Delta v=30\pm30$ km/s&Y \\
  small relative velocity& i.e. GalMer&&\\
  \\[.1 mm]
\enddata
\label{table:comparison}
\end{deluxetable*}

\section{Discussion}\label{sec:discussion}

In this section, we first synthesize the primary results of the preceeding
sections (summarized in Table~\ref{table:all}) and then consider the
implications for research in galaxy mergers and evolution.   

The J2049$-$0012 system is a pair of interacting galaxies. 
Its hallmark is an emission-line bridge that connects the two
galaxies, observed in recombination and forbidden line-emission. 
The observation of the bridge requires that the galaxies (now with a projected separation
of 38\,\hkpc) have had at least one tidal interaction.
One of the galaxies is currently in a quasar phase ($L_{\BOL} = 10^{45.7}$\ erg/s) and 
its ionizing radiation has
photoionized the gas in the bridge, producing the observed line-emission.
Using a simplistic volume model for the gas in the bridge, we 
constrained its surface and volume densities and provided an estimate
of the total gas mass ($\sim 10^8 M_\odot$). 

Despite its neighbor undergoing a violent quasar phase, the
companion galaxy shows no detectable AGN activity. Its photometry and spectrum
place it within or near the galaxy `green valley' and also
suggest a declining starburst stellar population (with burst age of
$300-800$~Myr). The companion galaxy has a current SFR 
of $\sim6 \ M_\odot \mbox{ yr}^{-1}$ and
a gas-phase metallicity consistent with solar. This metallicity is 
higher by $\approx 0.4$~dex than our
estimate for the bridge, as derived from
\textsc{Cloudy} photoionization modelling.

We made several mass estimates and constraints for both the quasar host 
and companion galaxy. In the context of galaxy mergers, the mass-ratio 
of the galaxies is a fundamental quantity because it strongly
influences the dynamics of the interaction.  
Our system has several pieces of evidence which suggest it has a mass
ratio of $\approx$ 1:4 and perhaps somewhat smaller.
Firstly, examination of the dynamical measurements
shows that the companion galaxy's rotation curve
flattens at a velocity of 110 \ \kms\ while the QSO host's
extends to a velocity of 200\ \kms\ and shows no signs of flattening.
Unless the companion galaxy
is seen relatively close to face on, it likely has a significantly
lower DM mass than the QSO host. 
Additionally, our estimates for the stellar masses of the two systems
is roughly 1:4.  
For the companion galaxy, we analyzed the SDSS
photometry to estimate a stellar mass of $1.9\times10^{10}M_\odot$. 
For the galaxy hosting the quasar, we cannot directly
measure the stellar light but instead estimate its stellar mass
using standard SMBH-host relations.
Taking the estimated SMBH mass of $10^{8.7}M_\odot$ \citep{shen08}
and the \cite{mag98} relation $M_{BH}\sim 0.006\ M_{host,\star}$, we
estimate $8\times10^{10} M_\odot$ for the stellar mass of the bulge component. 
This gives a stellar mass ratio of $\approx$1:4. If the quasar host galaxy has a
stellar mass in a disk component then this ratio shrinks even
smaller.  This result also hinges on whether the black hole currently
lies on the Magorrian relation.
We conclude, therefore, that the system has a mass ratio of approximately
1:4 with an uncertainty on the order of a factor of 2.  Therefore,
the event most likely either just satisfies the major-merger (1:4) definition or
lies somewhat below the canonical threshold.

Our observational analysis of the \qsonm\ system provides a
comprehensive and quantitative description of an interacting pair of
galaxies at $z \sim 0.4$.  In turn, it affords the opportunity to
directly test the standard paradigm for galaxy mergers as modeled by
numerical simulations \citep[e.g.][]{galmer, cox2006}.
The current theoretical understanding of galaxy interactions is that
galaxy mergers with mass ratios larger than $\approx$ 1:10 generally
follow a basic sequence of events with two possible phases of 
elevated star formation and AGN activity: one right after first passage
and the other during final coalescence. The absolute and relative strengths of
these two phases depends on the galaxy morphology (e.g.\ whether the
galaxies have gas, the presence of a bulge) and details of the
star-formation prescription.  
In such a paradigm, the combination of the detection of a tidal bridge and the current
projected separation of $38 h_{72}^{-1}$~kpc place a strong constraint on the merger stage.
Numerical simulations of major mergers 
with a variety of orbits \citep[e.g.][]{galmer} 
show that after first passage the galaxies move to a distance of
approximately 100 kpc where they reach their point of maximal separation (first apogee).
Subsequent passages and separations do not achieve separations
as large as 38~kpc. This result is nearly 
independent of gas fraction \citep{lotz_gass} and mass ratio as long
as the mass ratio is larger than 1:9 \citep{lotz_mass}. 
We conclude, therefore, that the \qsonm\ merger is after first passage
(the bridge requires at least one tidal interaction) and before second
passage.  

As described above, the current paradigm predicts that gas-rich
galaxies will undergo a burst of star-formation that initiates during
or just after first passage \citep[e.g.][]{mihos96}.  
This is also the only time aside from coalescence where gas inflow
is predicted to be sufficiently strong to inspire strong AGN activity
\citep{springel05, hop08}.   Regarding the \qsonm\ merger, we identify
examples of each:  one galaxy hosts a bright quasar and the companion
galaxy shows signatures of a starburst stellar
population (Figure~\ref{fig:ssp}).  
Consistent with theory of interactions causing tidal stripping and 
strong gaseous inflows which lead to large central star bursts,
the companion galaxy is observed to be consistent with being 
centrally concentrated (we measure a concentration of at least 2.61, 
limited by the image's spatial resolution). Such compact companions to quasars  
have been observed around other more nearby QSOs \citep{canalizo2001, stockton1982}. 
With standard stellar population modeling, we found the 
burst occurred $300-800$~Myr ago.  This timescale is significantly
shorter than the predicted period between first and second passage
for galaxies with our allowed range of mass ratios for
standard orbits \citep[$\approx 2$\,Gyr; e.g.][]{lotz_mass, galmer} .

In merger simulations, 
the level  of enhanced star formation during the first passage starburst is sensitive to
the star formation prescription \citep[see ][]{cox2006}.  Therefore characterizing this
phase may allow greater insights on sub-grid models. 
To this end, we would like to constrain the observed mass fraction of the burst. Unfortunately, various
degeneracies limit us to only constrain the burst to consist of between
$1-100\%$ of the stellar mass prior to the merger.

Another test of the merger paradigm combines the timing of the starburst and
the observed configuration of the merger. The current projected separation of
38 \hkpc\ combined with the timescale of $~500$ Myr requires that the galaxies
have been moving apart from each other at an average speed of approximately
$75$ \kms. This is consistent with the observed timescales and distances
seen in GalMer merger simulations \citep{galmer}.
The observed offset in radial velocities of the two galaxies also
follows the standard picture.
The centers of mass for the two
galaxies (as traced by the bright emission lines)
are moving at a relative radial velocity of $30\pm 30$\kms.
This is consistent with the system being close to apogee, as
predicted for a system with an age of $t \sim 500$\,Myr post first
passage \citep[e.g.][]{lotz_mass}. 

Lastly, merger simulations also offer robust predictions on the physical
characteristics of the tidal bridge. In the following few paragraphs, we discuss
how measurements of the bridge properties test
the merger paradigm and also present a means for gaining greater
understanding into specific mergers. 

Firstly, we note that the metallicity of the bridge, 
which we find to be a few tenths dex lower ($\Delta$[M/H]=$-0.36\pm0.23$)
than the companion galaxy, is consistent with having been tidally
stripped from the (presumed) lower metallicity outskirts of that galaxy. This is the standard
result of the merger paradigm first studied in detail by \cite{toomretoomre}. 
The outer parts of these extended tidal features arise from 
the outskirts of galactic disks because the tidal 
forces are greatest in these regions. 

Due to the unique emission line bridge of our system, we are able to measure the
gas surface density at a point midway between the two interacting galaxies ($\S$\ref{sec:column}).
Wishing to use this measurement to gain more insight on the interacting pair of galaxies,
we turn to numerical simulations. As this is not a standard observable and lacks sufficient 
literature detailing its evolution, we have explored
the GalMer database of galaxy mergers to characterize the
 bridge's evolution \citep{galmer}. Using a suite of 
1:1 and 1:10 galaxy merger simulations with a variety of spiral morphologies (1:1 
Sa-Sa, Sa-Sb, Sa-Sd, Sb-Sb, Sb-Sd, Sd-Sd ; 1:10 S0-Sa, S0-Sb, S0-Sd ),
we measured the gas surface density in the corresponding regions of these galaxy 
mergers at 64 viewing angles evenly spaced in solid angle at all applicable timesteps
(where the galaxies are separated by at least 20 projected kpc 
and have undergone at least 
one tidal interaction). The 1:1 mergers have 12 different orbits with 
each orbit also having
4 different relative orientations of the disks.  The 1:10 mergers are
limited 
by the availability of the models to only
have 6 orbits with only 1 generic disk orientation each.

We find that for 1:1 mergers, the bridge forms after first passage and
that its surface density 
reduces roughly exponentially with time 
(approximately $\propto \exp[-t/\tau]$ 
with $\tau\approx$100 Myr and the normalization correlating with the gas mass of the two
galaxies) as the material rains back onto
the two separating galaxies.  In the vast majority of cases, the gas density 
falls far below the measured surface density of our bridge\footnote{We note that while our
estimate of the column density is dependent on the filling factor ($\propto f^{1/2}$), the
dissipation is rapid enough that our result is relatively robust. We use our value of
the upper limit corresponding to $f=1$.}
before reaching second closest approach.
These findings indicate standard merger scenarios of galaxies
with high gas fraction can reproduce our observations provided we
have caught the system before maximal separation.

For 1:10 mergers we find that
 the gaseous bridge does not reach sufficiently high surface densities in the bridge
for any of the simulations we examined. We do note that in our 1:10 simulations the
larger galaxy does not contain any gas (we are limited by availability of simulations),
 which is likely a poor match to our system. 
This is unlikely to affect our overall results because in such a high mass ratio merger
 the contribution of the more massive galaxy to the extended tidal features will be minimal
 because the tidal force it feels is relatively small. In such a case, the smaller galaxy is
 much more greatly tidally perturbed, resulting in extended tidal
 features but at much smaller surface density than observed.  If the
 mass ratio of the \qsonm\ system were as small as 1:10, these models
 would be ruled out.

The kinematics of the bridge further enlighten us about the system 
(See Fig.~\ref{fig:kinematics}). The velocity centroids of the bridge as a
function of spatial position show that the side of the bridge closer to
the companion galaxy has a higher radial velocity than the side closer to
the quasar. If the gas is freely flowing towards the centers of mass
of the two galaxies, this suggests that the companion galaxy is background
to the quasar. This argument is confirmed by an examination of the GalMer merger simulations.
The measured velocity dispersion of $~50$ km/s indicates an intrinsic velocity 
dispersion of $~40$ km/s after we remove instrumental effects.
This is much higher than 
that found in 21 cm maps of tidal debris around local interacting galaxies
which have typical velocity dispersions of $~10-20$ km/s \citep{hibbard93, hibbard96}.
One possible explanation is suggested by the clumpy nature of the 
emission: portions of the bridge may have become
self-gravitating. Such structures are observed in the tidal features of
local galaxies and can have large velocity gradients $>40 $ km/s
\citep{weilbacher03, weilbacher02}. These structures are also 
a prediction of merger
simulations \citep{barnes}. It is possible that one or 
several of these structures could add to the velocity dispersion.
One immediate challenge to this possibility is the lack of obvious
star formation associated with the bridge. However,
if the bridge is fully ionized, it is possible that the large ionization 
fraction of the gas has suppressed star formation.
This is supported by the recent observational evidence that there is a 
neutral gas column density threshold
 ($\log N_{HI}>20.6\mbox{ cm}^{-2}$) for star formation in tidal material
\citep{maybhate}.  Another possible explanation is that dispersion has been
elevated by some sort of turbulence possibly associated with shocks that 
are thought to occur throughout the course of the merger as a result of the bulk motions
of the galaxies.

Given that we observe a bridge of material that we believe to be photoionized by the
quasar, one might ask if other tidal material should be observable (i.e. structures associated with
extended tidal tails). We have not detected such material, to
sensitive limits, in our Keck/LRIS longslit observations. There are
several reasons, however, why we may expect the non-detection 
of other extended emission. Firstly, there may be no tidal tails. 
\cite{roguesgallery} present a compilation 
of \ion{H}{1} maps of merging galaxies and find 16 examples 
of merging galaxy pairs (of 181 total unusual gas morphology examples) 
with an observed bridge and no tidal tails. 
Similarly, the GalMer merger simulation database \citep{galmer} has many examples of such systems.
Secondly, the tidal tails may simply not have fallen into the longslit
of our spectroscopic observations. 
Finally, the tidal material 
could have too low density to render it observable;
 emission is an $n^2$ process and a factor of 3 less dense gas results
 $\sim10$ times fainter emission. 
We conclude that the non-detection of additional, tidal material is
consistent with the observed bridge and standard merger scenarios.
Nevertheless, we are motivated to search for additional emission
beyond our longslit observation, e.g.\ with an IFU spectrometer.

Before concluding, we wish to consider a few issues related to the
`green valley' nature of the companion galaxy.  We remind the reader
that this is a rare and poorly understood phase
of galaxy evolution which nevertheless may be critical to
understanding the (expected) transition from the blue, star-forming
sequence to the `red and dead' sample. 
The first consideration
is to understand why the galaxy appears in this unusual portion of
color space. We have found that the peculiar colors are consistent
a declining starburst stellar population that would
dominate the SED for only another few hundred Myr. 
The next obvious consideration is whether 
the galaxy is currently moving toward the blue cloud or red sequence. Unfortunately, any such prediction hinges on
bold extrapolation of the future SFH. If we assume that the
observed SFR continues indefinitely (i.e.\ at $\approx 6 M_\odot$),
then we find that the galaxy evolves to becomes slightly bluer 
($u-g$ decreasing by 0.1  mag) 
over approximately 1 Gyr. However if the star formation ceases then the stellar populations will passively
evolve to become red after several hundred Myr. If the galaxies are to undergo further interactions with each other,
then the star formation history may follow one of the standard tracks towards quenching \citep[see discussion in][]{faber07}.

This system may also provide insights on some of the large number 
of newly discovered double peaked [\ion{O}{3}] emitters \citep{smith2010}. 
If observed with sufficiently
poor spatial resolution, it is possible that the bridge or the companion galaxy's emission associated with photoionization
from the quasar may result in the appearance of a double-peaked  [\ion{O}{3}]  profile separated by approximately 20 km/s resulting from just a single active SMBH. 
One could imagine that if caught closer in time to a close passage, where the radial velocity difference
of the two galaxies would be greater, the velocity difference may be large enough to produce the $\sim200$ km/s offsets observed in that sample.
This explanation is may be the most likely for a subset of
their systems (Rosario et al. in prep.). 

\section{Summary} 
In summary, we have 
compiled physical constraints on a galaxy merger at modest redshift
which allows for a direct test of the merger paradigm developed by
state-of-the-art simulations.
Impressively, the merger characteristics are consistent
with the standard paradigm for each of the comparisons summarized in Table~\ref{table:comparison}.
Our analysis presents a relatively clear-cut example of a quasar being triggered during a
first passage interaction. Observational characteristics allow it to be directly compared with simulations
to constrain triggering mechanisms.

We have copious amounts of information about the companion galaxy and
obvious questions emerge such as: Why is one galaxy in a quasar phase
while the other is not? Has the companion galaxy already undergone a
quasar phase? Are the black holes in each galaxy now and will they
later be on the many black hole-host relations? How does a first
passage quasar affect the evolution of a merger vs. one where neither
galaxy goes into a quasar phase until final coalescence? 
Paper~II presents the insights that we gain from this system about
some of these questions as well as discusses constraints on the
quasar's lifetime and isotropy that we get by examining the effects of
its ionizing radiation. 

Our work focuses on characterizing the system as well as possible.
A summary of our derived properties can be found in Table~\ref{table:all} and a
review of our conclusions follows below.
\begin{enumerate}
 \item We presented a unique interacting quasar/galaxy pair connected
   by a remarkable 38 \hkpc\ bridge of ionized gas visible in emission lines.
 \item While one galaxy is in a quasar phase, its companion shows no detectable AGN activity.
 \item After ruling out collisional ionization and radiative shocks,
   we conclude the bridge's source of ionization to be photoionization
   by the quasar light.
\item The companion galaxy's color and spectrum are consistent with a
  declining starburst stellar population. In this scenario, we date the burst stellar population to be
  $300-800$ Myr old. 
   This intermediate-age stellar population places the
  galaxy within the `green valley'.
\item We found the companion galaxy to have a SFR of $\sim6 \ M_\odot
  \mbox{ yr}^{-1}$ and to have a have a gas-phase oxygen abundance
  consistent with solar. 
 \item The bridge between the two galaxies was characterized to be
   dependent on a geometric parameter $\mathcal{D}$, which we constrain
   (see Fig.~\ref{fig:cartoon}). We found expressions for the volume
   density, column density along line of sight, and mass of the bridge
   as a function of this parameter.  
 \item With \textsc{cloudy} photoionization modeling, we determine the
   ionization parameter and metallicity in two separate patches of the
   bridge. The bridge is nearly fully ionized in hydrogen and is
   primiarly comprised of
   higher ionization species. The metallicity is consistent with being
   a few tenths dex smaller than the companion galaxy. 
 \item The bridge's velocity dispersion is larger than expected ($\sim40
   $km/s), possibly suggesting that some material in the bridge may be
   self-gravitating or there is some source of turbulence in the
   bridge. 
 \item Using the projected separation, we find the merger stage to be
   between first and second passage. This interpretation is consistent
   with the observed bridge and the inferred starburst stellar population. 
 \item Based on dynamical measurements and stellar photometry, we
   argue the mass ratio is likely $\sim$1:4. 
 \end{enumerate}
%
%

\acknowledgements  
Some of the data presented herein were obtained at the
W.M. Keck Observatory, which is operated as a scientific partnership
among the California Institute of Technology, the
University of California and the National Aeronautics and
Space Administration. The Observatory was made possible
by the generous financial support of the W.M. Keck Foundation.
The authors wish to recognize and acknowledge the very significant
cultural role and reverence that the summit of Mauna Kea has always
had within the indigenous Hawaiian community.  We are most fortunate
to have the opportunity to conduct observations from this mountain.
We thank the anonymous referee for comments which improved the content
and readability of this paper. 
R. L. dS. would like to thank T. J. Cox for helpful comments on 
the manuscript. R. L. dS. would also like to thank
Michele Fumagalli, Aaron Dutton, 
G\'{a}bor Worseck, Mark Krumholz, 
Greg Bryan, Phil Hopkins, Patrik Jonsson, Bill Matthews, 
Joe Miller, John O'Meara, Joel Primack, Enrico Ramirez-Ruiz, 
Greg Shields, and Marta Volonteri  for helpful discussions.
R.L.dS. and J.X.P. are partially supported by 
an NSF CAREER grant (AST-0548180). The work of R.L.dS. is 
supported under a National Science Foundation Graduate 
Research Fellowship.


\begin{thebibliography}{104}
\expandafter\ifx\csname natexlab\endcsname\relax\def\natexlab#1{#1}\fi

\bibitem[{{Abazajian} {et~al.}(2009){Abazajian}, {Adelman-McCarthy},
  {Ag{\"u}eros}, {Allam}, {Allende Prieto}, {An}, {Anderson}, {Anderson},
  {Annis}, {Bahcall}, {Bailer-Jones}, {Barentine}, {Bassett}, {Becker},
  {Beers}, {Bell}, {Belokurov}, {Berlind}, {Berman}, {Bernardi}, {Bickerton},
  {Bizyaev}, {Blakeslee}, {Blanton}, {Bochanski}, {Boroski}, {Brewington},
  {Brinchmann}, {Brinkmann}, {Brunner}, {Budav{\'a}ri}, {Carey}, {Carliles},
  {Carr}, {Castander}, {Cinabro}, {Connolly}, {Csabai}, {Cunha}, {Czarapata},
  {Davenport}, {de Haas}, {Dilday}, {Doi}, {Eisenstein}, {Evans}, {Evans},
  {Fan}, {Friedman}, {Frieman}, {Fukugita}, {G{\"a}nsicke}, {Gates},
  {Gillespie}, {Gilmore}, {Gonzalez}, {Gonzalez}, {Grebel}, {Gunn},
  {Gy{\"o}ry}, {Hall}, {Harding}, {Harris}, {Harvanek}, {Hawley}, {Hayes},
  {Heckman}, {Hendry}, {Hennessy}, {Hindsley}, {Hoblitt}, {Hogan}, {Hogg},
  {Holtzman}, {Hyde}, {Ichikawa}, {Ichikawa}, {Im}, {Ivezi{\'c}}, {Jester},
  {Jiang}, {Johnson}, {Jorgensen}, {Juri{\'c}}, {Kent}, {Kessler}, {Kleinman},
  {Knapp}, {Konishi}, {Kron}, {Krzesinski}, {Kuropatkin}, {Lampeitl},
  {Lebedeva}, {Lee}, {Lee}, {Leger}, {L{\'e}pine}, {Li}, {Lima}, {Lin}, {Long},
  {Loomis}, {Loveday}, {Lupton}, {Magnier}, {Malanushenko}, {Malanushenko},
  {Mandelbaum}, {Margon}, {Marriner}, {Mart{\'{\i}}nez-Delgado}, {Matsubara},
  {McGehee}, {McKay}, {Meiksin}, {Morrison}, {Mullally}, {Munn}, {Murphy},
  {Nash}, {Nebot}, {Neilsen}, {Newberg}, {Newman}, {Nichol}, {Nicinski},
  {Nieto-Santisteban}, {Nitta}, {Okamura}, {Oravetz}, {Ostriker}, {Owen},
  {Padmanabhan}, {Pan}, {Park}, {Pauls}, {Peoples}, {Percival}, {Pier}, {Pope},
  {Pourbaix}, {Price}, {Purger}, {Quinn}, {Raddick}, {Fiorentin}, {Richards},
  {Richmond}, {Riess}, {Rix}, {Rockosi}, {Sako}, {Schlegel}, {Schneider},
  {Scholz}, {Schreiber}, {Schwope}, {Seljak}, {Sesar}, {Sheldon}, {Shimasaku},
  {Sibley}, {Simmons}, {Sivarani}, {Smith}, {Smith}, {Smol{\v c}i{\'c}},
  {Snedden}, {Stebbins}, {Steinmetz}, {Stoughton}, {Strauss}, {Subba Rao},
  {Suto}, {Szalay}, {Szapudi}, {Szkody}, {Tanaka}, {Tegmark}, {Teodoro},
  {Thakar}, {Tremonti}, {Tucker}, {Uomoto}, {Vanden Berk}, {Vandenberg},
  {Vidrih}, {Vogeley}, {Voges}, {Vogt}, {Wadadekar}, {Watters}, {Weinberg},
  {West}, {White}, {Wilhite}, {Wonders}, {Yanny}, {Yocum}, {York}, {Zehavi},
  {Zibetti}, \& {Zucker}}]{sdssdr7}
{Abazajian}, K.~N., {et~al.} 2009, \apjs, 182, 543

\bibitem[{{Abel} \& {Satyapal}(2008)}]{nev_agn}
{Abel}, N.~P., \& {Satyapal}, S. 2008, \apj, 678, 686

\bibitem[{{Adelman-McCarthy} {et~al.}(2007){Adelman-McCarthy}, {Ag{\"u}eros},
  {Allam}, {Anderson}, {Anderson}, {Annis}, {Bahcall}, {Bailer-Jones},
  {Baldry}, {Barentine}, {Beers}, {Belokurov}, {Berlind}, {Bernardi},
  {Blanton}, {Bochanski}, {Boroski}, {Bramich}, {Brewington}, {Brinchmann},
  {Brinkmann}, {Brunner}, {Budav{\'a}ri}, {Carey}, {Carliles}, {Carr},
  {Castander}, {Connolly}, {Cool}, {Cunha}, {Csabai}, {Dalcanton}, {Doi},
  {Eisenstein}, {Evans}, {Evans}, {Fan}, {Finkbeiner}, {Friedman}, {Frieman},
  {Fukugita}, {Gillespie}, {Gilmore}, {Glazebrook}, {Gray}, {Grebel}, {Gunn},
  {de Haas}, {Hall}, {Harvanek}, {Hawley}, {Hayes}, {Heckman}, {Hendry},
  {Hennessy}, {Hindsley}, {Hirata}, {Hogan}, {Hogg}, {Holtzman}, {Ichikawa},
  {Ichikawa}, {Ivezi{\'c}}, {Jester}, {Johnston}, {Jorgensen}, {Juri{\'c}},
  {Kauffmann}, {Kent}, {Kleinman}, {Knapp}, {Kniazev}, {Kron}, {Krzesinski},
  {Kuropatkin}, {Lamb}, {Lampeitl}, {Lee}, {Leger}, {Lima}, {Lin}, {Long},
  {Loveday}, {Lupton}, {Mandelbaum}, {Margon}, {Mart{\'{\i}}nez-Delgado},
  {Matsubara}, {McGehee}, {McKay}, {Meiksin}, {Munn}, {Nakajima}, {Nash},
  {Neilsen}, {Newberg}, {Nichol}, {Nieto-Santisteban}, {Nitta}, {Oyaizu},
  {Okamura}, {Ostriker}, {Padmanabhan}, {Park}, {Peoples}, {Pier}, {Pope},
  {Pourbaix}, {Quinn}, {Raddick}, {Re Fiorentin}, {Richards}, {Richmond},
  {Rix}, {Rockosi}, {Schlegel}, {Schneider}, {Scranton}, {Seljak}, {Sheldon},
  {Shimasaku}, {Silvestri}, {Smith}, {Smol{\v c}i{\'c}}, {Snedden}, {Stebbins},
  {Stoughton}, {Strauss}, {SubbaRao}, {Suto}, {Szalay}, {Szapudi}, {Szkody},
  {Tegmark}, {Thakar}, {Tremonti}, {Tucker}, {Uomoto}, {Vanden Berk},
  {Vandenberg}, {Vidrih}, {Vogeley}, {Voges}, {Vogt}, {Weinberg}, {West},
  {White}, {Wilhite}, {Yanny}, {Yocum}, {York}, {Zehavi}, {Zibetti}, \&
  {Zucker}}]{sdssdr5}
{Adelman-McCarthy}, J.~K., {et~al.} 2007, \apjs, 172, 634

\bibitem[{{Allen} {et~al.}(2008){Allen}, {Groves}, {Dopita}, {Sutherland}, \&
  {Kewley}}]{shock_models}
{Allen}, M.~G., {Groves}, B.~A., {Dopita}, M.~A., {Sutherland}, R.~S., \&
  {Kewley}, L.~J. 2008, \apjs, 178, 20

\bibitem[Allende Prieto, Lambert, 
\& Asplund(2001)]{solarO1} Allende Prieto, C., Lambert, D.~L., \& Asplund, M.\ 2001, \apjl, 556, L63 

\bibitem[{{Arp}(1966)}]{arp}
{Arp}, H. 1966, {Atlas of peculiar galaxies}, ed. {Arp, H.}

\bibitem[Asplund et 
al.(2004)]{solarO2} Asplund, M., Grevesse, N., Sauval, A.~J., Allende Prieto, C., \& Kiselman, D.\ 2004, \aap, 417, 751 

\bibitem[{{Baldwin} {et~al.}(1981){Baldwin}, {Phillips}, \& {Terlevich}}]{bpt}
{Baldwin}, J.~A., {Phillips}, M.~M., \& {Terlevich}, R. 1981, \pasp, 93, 5

\bibitem[{{Barnes} \& {Hernquist}(1992)}]{barnes}
{Barnes}, J.~E., \& {Hernquist}, L. 1992, \nat, 360, 715

\bibitem[{{Bennert} {et~al.}(2008){Bennert}, {Canalizo}, {Jungwiert},
  {Stockton}, {Schweizer}, {Peng}, \& {Lacy}}]{ben08}
{Bennert}, N., {Canalizo}, G., {Jungwiert}, B., {Stockton}, A., {Schweizer},
  F., {Peng}, C.~Y., \& {Lacy}, M. 2008, \apj, 677, 846

\bibitem[{{Blanton} \& {Roweis}(2007)}]{kcorrect}
{Blanton}, M.~R., \& {Roweis}, S. 2007, \aj, 133, 734

\bibitem[{{Bohlin}(1996)}]{bohlin96}
{Bohlin}, R.~C. 1996, \aj, 111, 1743

\bibitem[{{Bohlin} {et~al.}(2001){Bohlin}, {Dickinson}, \&
  {Calzetti}}]{bohlin01}
{Bohlin}, R.~C., {Dickinson}, M.~E., \& {Calzetti}, D. 2001, \aj, 122, 2118

\bibitem[{{Booth} \& {Schaye}(2009)}]{booth2009}
{Booth}, C.~M., \& {Schaye}, J. 2009, \mnras, 398, 53

\bibitem[{{Brammer} {et~al.}(2009){Brammer}, {Whitaker}, {van Dokkum},
  {Marchesini}, {Labb{\'e}}, {Franx}, {Kriek}, {Quadri}, {Illingworth}, {Lee},
  {Muzzin}, \& {Rudnick}}]{bram09}
{Brammer}, G.~B., {et~al.} 2009, \apjl, 706, L173

\bibitem[{{Bruzual} \& {Charlot}(2003)}]{bc03}
{Bruzual}, G., \& {Charlot}, S. 2003, \mnras, 344, 1000

\bibitem[{{Bullock} {et~al.}(2001){Bullock}, {Kolatt}, {Sigad}, {Somerville},
  {Kravtsov}, {Klypin}, {Primack}, \& {Dekel}}]{Bullock2001}
{Bullock}, J.~S., {Kolatt}, T.~S., {Sigad}, Y., {Somerville}, R.~S.,
  {Kravtsov}, A.~V., {Klypin}, A.~A., {Primack}, J.~R., \& {Dekel}, A. 2001,
  \mnras, 321, 559

\bibitem[{{Canalizo} \& {Stockton}(2001)}]{canalizo2001}
{Canalizo}, G., \& {Stockton}, A. 2001, \apj, 555, 719

\bibitem[{{Cardelli} {et~al.}(1989){Cardelli}, {Clayton}, \&
  {Mathis}}]{cardelli_dust}
{Cardelli}, J.~A., {Clayton}, G.~C., \& {Mathis}, J.~S. 1989, \apj, 345, 245

\bibitem[{{Chabrier}(2003)}]{chab}
{Chabrier}, G. 2003, \pasp, 115, 763

\bibitem[{{Charlot} \& {Fall}(2000)}]{charlot_fall2000}
{Charlot}, S., \& {Fall}, S.~M. 2000, \apj, 539, 718

\bibitem[Chilingarian et 
al.(2010)]{galmer} Chilingarian, I.~V., Di Matteo, 
P., Combes, F., Melchior, A.-L., \& Semelin, B.\ 2010, \aap, 518, A61 

\bibitem[{{Coil} {et~al.}(2004){Coil}, {Newman}, {Kaiser}, {Davis}, {Ma},
  {Kocevski}, \& {Koo}}]{deep3}
{Coil}, A.~L., {Newman}, J.~A., {Kaiser}, N., {Davis}, M., {Ma}, C.,
  {Kocevski}, D.~D., \& {Koo}, D.~C. 2004, \apj, 617, 765

\bibitem[{{Conroy} \& {Wechsler}(2009)}]{conroy2009}
{Conroy}, C., \& {Wechsler}, R.~H. 2009, \apj, 696, 620

\bibitem[{{Conselice}(2003)}]{conselice2003}
{Conselice}, C.~J. 2003, \apjs, 147, 1

\bibitem[{{Cox} {et~al.}(2006){Cox}, {Jonsson}, {Primack}, \&
  {Somerville}}]{cox2006}
{Cox}, T.~J., {Jonsson}, P., {Primack}, J.~R., \& {Somerville}, R.~S. 2006,
  \mnras, 373, 1013

\bibitem[{{Croom} {et~al.}(2009){Croom}, {Richards}, {Shanks}, {Boyle},
  {Strauss}, {Myers}, {Nichol}, {Pimbblet}, {Ross}, {Schneider}, {Sharp}, \&
  {Wake}}]{qsolumfcn}
{Croom}, S.~M., {et~al.} 2009, \mnras, 399, 1755

\bibitem[{{Cutri} {et~al.}(2003){Cutri}, {Skrutskie}, {van Dyk}, {Beichman},
  {Carpenter}, {Chester}, {Cambresy}, {Evans}, {Fowler}, {Gizis}, {Howard},
  {Huchra}, {Jarrett}, {Kopan}, {Kirkpatrick}, {Light}, {Marsh}, {McCallon},
  {Schneider}, {Stiening}, {Sykes}, {Weinberg}, {Wheaton}, {Wheelock}, \&
  {Zacarias}}]{2mass}
{Cutri}, R.~M., {et~al.} 2003, {2MASS All Sky Catalog of point sources.}, ed.
  {Cutri, R.~M., Skrutskie, M.~F., van Dyk, S., Beichman, C.~A., Carpenter,
  J.~M., Chester, T., Cambresy, L., Evans, T., Fowler, J., Gizis, J., Howard,
  E., Huchra, J., Jarrett, T., Kopan, E.~L., Kirkpatrick, J.~D., Light, R.~M.,
  Marsh, K.~A., McCallon, H., Schneider, S., Stiening, R., Sykes, M., Weinberg,
  M., Wheaton, W.~A., Wheelock, S., \& Zacarias, N.}

\bibitem[{{Davis} {et~al.}(2003){Davis}, {Faber}, {Newman}, {Phillips},
  {Ellis}, {Steidel}, {Conselice}, {Coil}, {Finkbeiner}, {Koo}, {Guhathakurta},
  {Weiner}, {Schiavon}, {Willmer}, {Kaiser}, {Luppino}, {Wirth}, {Connolly},
  {Eisenhardt}, {Cooper}, \& {Gerke}}]{deep1}
{Davis}, M., {et~al.} 2003, in Presented at the Society of Photo-Optical
  Instrumentation Engineers (SPIE) Conference, Vol. 4834, Society of
  Photo-Optical Instrumentation Engineers (SPIE) Conference Series, ed.
  {P.~Guhathakurta}, 161--172

\bibitem[{{Davis} {et~al.}(2007){Davis}, {Guhathakurta}, {Konidaris}, {Newman},
  {Ashby}, {Biggs}, {Barmby}, {Bundy}, {Chapman}, {Coil}, {Conselice},
  {Cooper}, {Croton}, {Eisenhardt}, {Ellis}, {Faber}, {Fang}, {Fazio},
  {Georgakakis}, {Gerke}, {Goss}, {Gwyn}, {Harker}, {Hopkins}, {Huang},
  {Ivison}, {Kassin}, {Kirby}, {Koekemoer}, {Koo}, {Laird}, {Le Floc'h}, {Lin},
  {Lotz}, {Marshall}, {Martin}, {Metevier}, {Moustakas}, {Nandra}, {Noeske},
  {Papovich}, {Phillips}, {Rich}, {Rieke}, {Rigopoulou}, {Salim},
  {Schiminovich}, {Simard}, {Smail}, {Small}, {Weiner}, {Willmer}, {Willner},
  {Wilson}, {Wright}, \& {Yan}}]{deep2}
{Davis}, M., {et~al.} 2007, \apjl, 660, L1

\bibitem[Dekel 
\& Birnboim(2006)]{dekel} Dekel, A., \& Birnboim, Y.\ 2006, \mnras, 368, 2 


\bibitem[{{Di Matteo} {et~al.}(2007){Di Matteo}, {Combes}, {Melchior}, \&
  {Semelin}}]{dimatteo2007}
{Di Matteo}, P., {Combes}, F., {Melchior}, A., \& {Semelin}, B. 2007, \aap,
  468, 61

\bibitem[{{Dressler} \& {Gunn}(1983)}]{dress83}
{Dressler}, A., \& {Gunn}, J.~E. 1983, \apj, 270, 7

\bibitem[{Ellison}{et~al.} (2008){Ellison},{Patton},{Simard}, \&{McConnachie}]{ellison} {Ellison}, S.~L., {Patton}, D.~R., {Simard}, L., \& {McConnachie}, A.~W.\ 2008, \aj, 135, 1877 

\bibitem[{{Faber} {et~al.}(2007){Faber}, {Willmer}, {Wolf}, {Koo}, {Weiner},
  {Newman}, {Im}, {Coil}, {Conroy}, {Cooper}, {Davis}, {Finkbeiner}, {Gerke},
  {Gebhardt}, {Groth}, {Guhathakurta}, {Harker}, {Kaiser}, {Kassin},
  {Kleinheinrich}, {Konidaris}, {Kron}, {Lin}, {Luppino}, {Madgwick},
  {Meisenheimer}, {Noeske}, {Phillips}, {Sarajedini}, {Schiavon}, {Simard},
  {Szalay}, {Vogt}, \& {Yan}}]{faber07}
{Faber}, S.~M., {et~al.} 2007, \apj, 665, 265

\bibitem[{{Fakhouri} {et~al.}(2010){Fakhouri}, {Ma}, \&
  {Boylan-Kolchin}}]{mergertrees}
{Fakhouri}, O., {Ma}, C., \& {Boylan-Kolchin}, M. 2010, \mnras, 406, 2267

\bibitem[{{Ferland} {et~al.}(1998){Ferland}, {Korista}, {Verner}, {Ferguson},
  {Kingdon}, \& {Verner}}]{cloudy}
{Ferland}, G.~J., {Korista}, K.~T., {Verner}, D.~A., {Ferguson}, J.~W.,
  {Kingdon}, J.~B., \& {Verner}, E.~M. 1998, \pasp, 110, 761

\bibitem[{{Ferrarese}(2002)}]{beyondbulge}
{Ferrarese}, L. 2002, \apj, 578, 90

\bibitem[{{Ferrarese} \& {Merritt}(2000)}]{ferrarese2000}
{Ferrarese}, L., \& {Merritt}, D. 2000, \apjl, 539, L9

\bibitem[{{Fontana} {et~al.}(2006){Fontana}, {Salimbeni}, {Grazian},
  {Giallongo}, {Pentericci}, {Nonino}, {Fontanot}, {Menci}, {Monaco},
  {Cristiani}, {Vanzella}, {de Santis}, \& {Gallozzi}}]{fontana2006}
{Fontana}, A., {et~al.} 2006, \aap, 459, 745

\bibitem[{{Fu} \& {Stockton}(2009)}]{fu09}
{Fu}, H., \& {Stockton}, A. 2009, \apj, 690, 953

\bibitem[{{Gebhardt} {et~al.}(2000){Gebhardt}, {Bender}, {Bower}, {Dressler},
  {Faber}, {Filippenko}, {Green}, {Grillmair}, {Ho}, {Kormendy}, {Lauer},
  {Magorrian}, {Pinkney}, {Richstone}, \& {Tremaine}}]{geb2000}
{Gebhardt}, K., {et~al.} 2000, \apjl, 539, L13

\bibitem[{{Graham} {et~al.}(2001){Graham}, {Erwin}, {Caon}, \&
  {Trujillo}}]{graham2001}
{Graham}, A.~W., {Erwin}, P., {Caon}, N., \& {Trujillo}, I. 2001, \apjl, 563,
  L11

\bibitem[{{Green} {et~al.}(2010){Green}, {Myers}, {Barkhouse}, {Mulchaey},
  {Bennert}, {Cox}, \& {Aldcroft}}]{green2010}
{Green}, P.~J., {Myers}, A.~D., {Barkhouse}, W.~A., {Mulchaey}, J.~S.,
  {Bennert}, V.~N., {Cox}, T.~J., \& {Aldcroft}, T.~L. 2010, \apj, 710, 1578

\bibitem[{{Greene} {et~al.}(2009){Greene}, {Zakamska}, {Liu}, {Barth}, \&
  {Ho}}]{gre09}
{Greene}, J.~E., {Zakamska}, N.~L., {Liu}, X., {Barth}, A.~J., \& {Ho}, L.~C.
  2009, \apj, 702, 441

\bibitem[{{Hibbard} {et~al.}(2001){Hibbard}, {van der Hulst}, {Barnes}, \&
  {Rich}}]{hib01}
{Hibbard}, J.~E., {van der Hulst}, J.~M., {Barnes}, J.~E., \& {Rich}, R.~M.
  2001, \aj, 122, 2969

\bibitem[{{Hibbard} \& {van Gorkom}(1993)}]{hibbard93}
{Hibbard}, J.~E., \& {van Gorkom}, J.~H. 1993, in Astronomical Society of the
  Pacific Conference Series, Vol.~48, The Globular Cluster-Galaxy Connection,
  ed. {G.~H.~Smith \& J.~P.~Brodie}, 619--+

\bibitem[{{Hibbard} \& {van Gorkom}(1996)}]{hibbard96}
{Hibbard}, J.~E., \& {van Gorkom}, J.~H. 1996, \aj, 111, 655

\bibitem[{{Hibbard} {et~al.}(2001){Hibbard}, {van Gorkom}, {Rupen}, {Schiminovich}}]{roguesgallery} Hibbard, J.~E., van Gorkom, J.~H., Rupen, M.~P., \& Schiminovich, D.\ 2001, Gas and Galaxy Evolution, 240, 657 

\bibitem[{{Ho}(2005)}]{ho2005}
{Ho}, L.~C. 2005, \apj, 629, 680

\bibitem[{{Hopkins} {et~al.}(2008{\natexlab{a}}){Hopkins}, {Hernquist}, {Cox},
  \& {Kere{\v s}}}]{hop08}
{Hopkins}, P.~F., {Hernquist}, L., {Cox}, T.~J., \& {Kere{\v s}}, D.
  2008{\natexlab{a}}, \apjs, 175, 356

\bibitem[{{Hopkins} {et~al.}(2008{\natexlab{b}}){Hopkins}, {Hernquist}, {Cox},
  \& {Kere{\v s}}}]{hop08framework1}
---. 2008{\natexlab{b}}, \apjs, 175, 356

\bibitem[{{Kauffmann} {et~al.}(2003{\natexlab{a}}){Kauffmann}, {Heckman},
  {White}, {Charlot}, {Tremonti}, {Peng}, {Seibert}, {Brinkmann}, {Nichol},
  {SubbaRao}, \& {York}}]{kauffmann2003}
{Kauffmann}, G., {et~al.} 2003{\natexlab{a}}, \mnras, 341, 54

\bibitem[{{Kauffmann} {et~al.}(2003{\natexlab{b}}){Kauffmann}, {Heckman},
  {Tremonti}, {Brinchmann}, {Charlot}, {White}, {Ridgway}, {Brinkmann},
  {Fukugita}, {Hall}, {Ivezi{\'c}}, {Richards}, \& {Schneider}}]{kauffbpt}
---. 2003{\natexlab{b}}, \mnras, 346, 1055

\bibitem[{{Kelly} {et~al.}(2010){Kelly}, {Vestergaard}, {Fan}, {Hopkins},
  {Hernquist}, \& {Siemiginowska}}]{blqso_massfcn}
{Kelly}, B.~C., {Vestergaard}, M., {Fan}, X., {Hopkins}, P., {Hernquist}, L.,
  \& {Siemiginowska}, A. 2010, \apj, 719, 1315

\bibitem[{{Kennicutt}(1998)}]{kennicutt98}
{Kennicutt}, Jr., R.~C. 1998, \araa, 36, 189

\bibitem[{{Kewley} {et~al.}(2001){Kewley}, {Dopita}, {Sutherland}, {Heisler},
  \& {Trevena}}]{kewl01}
{Kewley}, L.~J., {Dopita}, M.~A., {Sutherland}, R.~S., {Heisler}, C.~A., \&
  {Trevena}, J. 2001, \apj, 556, 121

\bibitem[{{Kocevski} {et~al.}(2009){Kocevski}, {Lubin}, {Lemaux}, {Gal},
  {Fassnacht}, {Lin}, \& {Squires}}]{kocevski}
{Kocevski}, D.~D., {Lubin}, L.~M., {Lemaux}, B.~C., {Gal}, R.~R., {Fassnacht},
  C.~D., {Lin}, R., \& {Squires}, G.~K. 2009, \apj, 700, 901

\bibitem[{{Kong} {et~al.}(2006){Kong}, {Wu}, {Wang}, \& {Han}}]{kong06}
{Kong}, M., {Wu}, X., {Wang}, R., \& {Han}, J. 2006, Chinese Journal of
  Astronomy and Astrophysics, 6, 396

\bibitem[{{Lotz} {et~al.}(2008{\natexlab{a}}){Lotz}, {Jonsson}, {Cox}, \&
  {Primack}}]{lotz2008}
{Lotz}, J.~M., {Jonsson}, P., {Cox}, T.~J., \& {Primack}, J.~R.
  2008{\natexlab{a}}, \mnras, 391, 1137

\bibitem[{{Lotz} {et~al.}(2010{\natexlab{a}}){Lotz}, {Jonsson}, {Cox}, \&
  {Primack}}]{lotz_gass}
---. 2010{\natexlab{a}}, \mnras, 404, 590

\bibitem[{{Lotz} {et~al.}(2010{\natexlab{b}}){Lotz}, {Jonsson}, {Cox}, \&
  {Primack}}]{lotz_mass}
---. 2010{\natexlab{b}}, \mnras, 404, 575

\bibitem[{{Lotz} {et~al.}(2008{\natexlab{b}}){Lotz}, {Davis}, {Faber},
  {Guhathakurta}, {Gwyn}, {Huang}, {Koo}, {Le Floc'h}, {Lin}, {Newman},
  {Noeske}, {Papovich}, {Willmer}, {Coil}, {Conselice}, {Cooper}, {Hopkins},
  {Metevier}, {Primack}, {Rieke}, \& {Weiner}}]{lotz08}
{Lotz}, J.~M., {et~al.} 2008{\natexlab{b}}, \apj, 672, 177

\bibitem[{{Magorrian} {et~al.}(1998){Magorrian}, {Tremaine}, {Richstone},
  {Bender}, {Bower}, {Dressler}, {Faber}, {Gebhardt}, {Green}, {Grillmair},
  {Kormendy}, \& {Lauer}}]{mag98}
{Magorrian}, J., {et~al.} 1998, \aj, 115, 2285

\bibitem[{{Martin} {et~al.}(2005){Martin}, {Fanson}, {Schiminovich},
  {Morrissey}, {Friedman}, {Barlow}, {Conrow}, {Grange}, {Jelinsky},
  {Milliard}, {Siegmund}, {Bianchi}, {Byun}, {Donas}, {Forster}, {Heckman},
  {Lee}, {Madore}, {Malina}, {Neff}, {Rich}, {Small}, {Surber}, {Szalay},
  {Welsh}, \& {Wyder}}]{galex}
{Martin}, D.~C., {et~al.} 2005, \apjl, 619, L1


\bibitem[Martin et al.(2007)]{martin2007} Martin, D.~C., et al.\ 
2007, \apjs, 173, 342 

\bibitem[{{Martini} \& {Schneider}(2003)}]{mart03}
{Martini}, P., \& {Schneider}, D.~P. 2003, \apjl, 597, L109

\bibitem[{{Martini} \& {Weinberg}(2001)}]{mart01}
{Martini}, P., \& {Weinberg}, D.~H. 2001, \apj, 547, 12

\bibitem[{{Matthews} \& {Sandage}(1963)}]{mat63}
{Matthews}, T.~A., \& {Sandage}, A.~R. 1963, \apj, 138, 30

\bibitem[{{Maybhate} {et~al.}(2007){Maybhate}, {Masiero}, {Hibbard},
  {Charlton}, {Palma}, {Knierman}, \& {English}}]{maybhate}
{Maybhate}, A., {Masiero}, J., {Hibbard}, J.~E., {Charlton}, J.~C., {Palma},
  C., {Knierman}, K.~A., \& {English}, J. 2007, \mnras, 381, 59

\bibitem[{{Mihos} \& {Hernquist}(1994)}]{mihos1994b}
{Mihos}, J.~C., \& {Hernquist}, L. 1994, \apjl, 431, L9

\bibitem[{{Mihos} \& {Hernquist}(1996)}]{mihos96}
---. 1996, \apj, 464, 641

\bibitem[{{Miller} \& {Stone}(1993)}]{kast}
Miller, J. S., \& Stone, R. P. S. 1993, Lick Observatory Technical Reports 66 (Santa Cruz, CA; Lick Observatory)

\bibitem[{{Nandra} {et~al.}(2007){Nandra}, {Georgakakis}, {Willmer}, {Cooper},
  {Croton}, {Davis}, {Faber}, {Koo}, {Laird}, \& {Newman}}]{nandra07}
{Nandra}, K., {et~al.} 2007, \apjl, 660, L11

\bibitem[Noeske et al.(2007)]{noeske07} Noeske, K.~G., et al.\ 
2007, \apjl, 660, L47 

\bibitem[Noeske et al.(2007)]{noeske07_2} Noeske, K.~G., et al.\ 
2007, \apjl, 660, L43 

\bibitem[{{Oke} {et~al.}(1995){Oke}, {Cohen}, {Carr}, {Cromer}, {Dingizian},
  {Harris}, {Labrecque}, {Lucinio}, {Schaal}, {Epps}, \& {Miller}}]{lris}
{Oke}, J.~B., {et~al.} 1995, \pasp, 107, 375

\bibitem[{{Osterbrock} \& {Ferland}(2006)}]{agn2}
{Osterbrock}, D.~E., \& {Ferland}, G.~J. 2006, {Astrophysics of gaseous nebulae
  and active galactic nuclei}, ed. {Osterbrock, D.~E.~\& Ferland, G.~J.}

\bibitem[{{Oyaizu} {et~al.}(2008){Oyaizu}, {Lima}, {Cunha}, {Lin}, {Frieman},
  \& {Sheldon}}]{sdssphotz}
{Oyaizu}, H., {Lima}, M., {Cunha}, C.~E., {Lin}, H., {Frieman}, J., \&
  {Sheldon}, E.~S. 2008, \apj, 674, 768

\bibitem[{{Penston} {et~al.}(1990){Penston}, {Robinson}, {Alloin},
  {Appenzeller}, {Aretxaga}, {Axon}, {Baribaud}, {Barthel}, {Baum}, {Boisson},
  {de Bruyn}, {Clavel}, {Colina}, {Dennefeld}, {Diaz}, {Dietrich}, {Durret},
  {Dyson}, {Gondhalekar}, {van Groningen}, {Jablonka}, {Jackson},
  {Kollatschny}, {Laurikainen}, {Lawrence}, {Masegosa}, {McHardy}, {Meurs},
  {Miley}, {Moles}, {O'Brien}, {O'Dea}, {del Olmo}, {Pedlar}, {Perea}, {Perez},
  {Perez-Fournon}, {Perry}, {Pilbratt}, {Rees}, {Robson}, {Rodriguez-Pascual},
  {Rodriguez Espinosa}, {Santos-Lleo}, {Schilizzi}, {Stasi{\'n}ska}, {Stirpe},
  {Tadhunter}, {Terlevich}, {Terlevich}, {Unger}, {Vila-Vilaro}, {Vilchez},
  {Wagner}, {Ward}, \& {Yates}}]{pens90}
{Penston}, M.~V., {et~al.} 1990, \aap, 236, 53

\bibitem[{{Pettini} \& {Pagel}(2004)}]{pp04}
{Pettini}, M., \& {Pagel}, B.~E.~J. 2004, \mnras, 348, L59

\bibitem[{{Quintero} {et~al.}(2004){Quintero}, {Hogg}, {Blanton}, {Schlegel},
  {Eisenstein}, {Gunn}, {Brinkmann}, {Fukugita}, {Glazebrook}, \&
  {Goto}}]{quint04}
{Quintero}, A.~D., {et~al.} 2004, \apj, 602, 190

\bibitem[Reichard et al.(2009)]{reichard} Reichard, T.~A., 
Heckman, T.~M., Rudnick, G., Brinchmann, J., Kauffmann, G., 
\& Wild, V.\ 2009, \apj, 691, 1005 

\bibitem[{{Richards} {et~al.}(2002){Richards}, {Vanden Berk}, {Reichard},
  {Hall}, {Schneider}, {SubbaRao}, {Thakar}, \& {York}}]{richards2002}
{Richards}, G.~T., {Vanden Berk}, D.~E., {Reichard}, T.~A., {Hall}, P.~B.,
  {Schneider}, D.~P., {SubbaRao}, M., {Thakar}, A.~R., \& {York}, D.~G. 2002,
  \aj, 124, 1

\bibitem[{{Rosario} {et~al.}(2010){Rosario}, {Shields}, {Taylor}, {Salviander},
  \& {Smith}}]{rosario10}
{Rosario}, D.~J., {Shields}, G.~A., {Taylor}, G.~B., {Salviander}, S., \&
  {Smith}, K.~L. 2010, \apj, 716, 131

\bibitem[{{Rubin} {et~al.}(2010){Rubin}, {Weiner}, {Koo}, {Martin},
  {Prochaska}, {Coil}, \& {Newman}}]{rubin2010}
{Rubin}, K.~H.~R., {Weiner}, B.~J., {Koo}, D.~C., {Martin}, C.~L., {Prochaska},
  J.~X., {Coil}, A.~L., \& {Newman}, J.~A. 2010, \apj, 719, 1503

\bibitem[{{Schawinski} {et~al.}(2009){Schawinski}, {Virani}, {Simmons}, {Urry},
  {Treister}, {Kaviraj}, \& {Kushkuley}}]{schaw09agngreenval}
{Schawinski}, K., {Virani}, S., {Simmons}, B., {Urry}, C.~M., {Treister}, E.,
  {Kaviraj}, S., \& {Kushkuley}, B. 2009, \apjl, 692, L19

\bibitem[{{Schlegel} {et~al.}(1998){Schlegel}, {Finkbeiner}, \&
  {Davis}}]{schlegel}
{Schlegel}, D.~J., {Finkbeiner}, D.~P., \& {Davis}, M. 1998, \apj, 500, 525

\bibitem[{{Shen} {et~al.}(2008){Shen}, {Greene}, {Strauss}, {Richards}, \&
  {Schneider}}]{shen08}
{Shen}, Y., {Greene}, J.~E., {Strauss}, M.~A., {Richards}, G.~T., \&
  {Schneider}, D.~P. 2008, \apj, 680, 169

\bibitem[{{Smith} {et~al.}(2010){Smith}, {Shields}, {Bonning}, {McMullen},
  {Rosario}, \& {Salviander}}]{smith2010}
{Smith}, K.~L., {Shields}, G.~A., {Bonning}, E.~W., {McMullen}, C.~C.,
  {Rosario}, D.~J., \& {Salviander}, S. 2010, \apj, 716, 866

\bibitem[{{Soltan}(1982)}]{soltan}
{Soltan}, A. 1982, \mnras, 200, 115

\bibitem[{{Springel} {et~al.}(2005{\natexlab{a}}){Springel}, {Di Matteo}, \&
  {Hernquist}}]{springel05}
{Springel}, V., {Di Matteo}, T., \& {Hernquist}, L. 2005{\natexlab{a}}, \mnras,
  361, 776

\bibitem[{{Springel} {et~al.}(2005{\natexlab{b}}){Springel}, {White},
  {Jenkins}, {Frenk}, {Yoshida}, {Gao}, {Navarro}, {Thacker}, {Croton},
  {Helly}, {Peacock}, {Cole}, {Thomas}, {Couchman}, {Evrard}, {Colberg}, \&
  {Pearce}}]{millennium}
{Springel}, V., {et~al.} 2005{\natexlab{b}}, \nat, 435, 629

\bibitem[{{Stockton}(1982)}]{stockton1982}
{Stockton}, A. 1982, \apj, 257, 33

\bibitem[{{Stockton} \& {MacKenty}(1987)}]{stock87}
{Stockton}, A., \& {MacKenty}, J.~W. 1987, \apj, 316, 584

\bibitem[{{Sutter} \& {Ricker}(2010)}]{sutter2010}
{Sutter}, P.~M., \& {Ricker}, P.~M. 2010, ArXiv e-prints

\bibitem[{{Szentgyorgyi} {et~al.}(2000){Szentgyorgyi}, {Raymond}, {Hester}, \&
  {Curiel}}]{neV}
{Szentgyorgyi}, A.~H., {Raymond}, J.~C., {Hester}, J.~J., \& {Curiel}, S. 2000,
  \apj, 529, 279

\bibitem[{{Tadhunter}(1996)}]{tad96}
{Tadhunter}, C. 1996, {Optical spectroscopy of Cygnus A: mixed evidence for a
  hidden quasar}, ed. {Carilli, C.~L.~\& Harris, D.~E.}, 33--+

\bibitem[{{Toomre}(1977)}]{toomre77}
{Toomre}, A. 1977, in Evolution of Galaxies and Stellar Populations, ed.
  {B.~M.~Tinsley \& R.~B.~Larson}, 401--+

\bibitem[{{Toomre} \& {Toomre}(1972{\natexlab{a}})}]{toomre72}
{Toomre}, A., \& {Toomre}, J. 1972{\natexlab{a}}, \apj, 178, 623

\bibitem[{{Toomre} \& {Toomre}(1972{\natexlab{b}})}]{toomretoomre}
---. 1972{\natexlab{b}}, \apj, 178, 623

\bibitem[{{Trammell} {et~al.}(2007){Trammell}, {Vanden Berk}, {Schneider},
  {Richards}, {Hall}, {Anderson}, \& {Brinkmann}}]{tram07}
{Trammell}, G.~B., {Vanden Berk}, D.~E., {Schneider}, D.~P., {Richards}, G.~T.,
  {Hall}, P.~B., {Anderson}, S.~F., \& {Brinkmann}, J. 2007, \aj, 133, 1780

\bibitem[{{Veilleux} \& {Osterbrock}(1987)}]{veil87}
{Veilleux}, S., \& {Osterbrock}, D.~E. 1987, \apjs, 63, 295

\bibitem[{{Vergani} {et~al.}(2010){Vergani}, {Zamorani}, {Lilly}, {Lamareille},
  {Halliday}, {Scodeggio}, {Vignali}, {Ciliegi}, {Bolzonella}, {Bondi},
  {Kova{\v c}}, {Knobel}, {Zucca}, {Caputi}, {Pozzetti}, {Bardelli}, {Mignoli},
  {Iovino}, {Carollo}, {Contini}, {Kneib}, {Le F{\`e}vre}, {Mainieri},
  {Renzini}, {Bongiorno}, {Coppa}, {Cucciati}, {de la Torre}, {de Ravel},
  {Franzetti}, {Garilli}, {Kampczyk}, {Le Borgne}, {Le Brun}, {Maier}, {Pello},
  {Peng}, {Perez Montero}, {Ricciardelli}, {Silverman}, {Tanaka}, {Tasca},
  {Tresse}, {Abbas}, {Bottini}, {Cappi}, {Cassata}, {Cimatti}, {Guzzo},
  {Koekemoer}, {Leauthaud}, {Maccagni}, {Marinoni}, {McCracken}, {Memeo},
  {Meneux}, {Oesch}, {Porciani}, {Scaramella}, {Capak}, {Sanders}, {Scoville},
  \& {Taniguchi}}]{vergani}
{Vergani}, D., {et~al.} 2010, \aap, 509, A42+

\bibitem[{{Vestergaard} \& {Peterson}(2006)}]{vest06}
{Vestergaard}, M., \& {Peterson}, B.~M. 2006, \apj, 641, 689

\bibitem[{{Villar-Mart{\'{\i}}n} {et~al.}(2010){Villar-Mart{\'{\i}}n},
  {Tadhunter}, {P{\'e}rez}, {Humphrey}, {Mart{\'{\i}}nez-Sansigre},
  {Gonz{\'a}lez Delgado}, \& {P{\'e}rez-Torres}}]{villarmartin10}
{Villar-Mart{\'{\i}}n}, M., {Tadhunter}, C., {P{\'e}rez}, E., {Humphrey}, A.,
  {Mart{\'{\i}}nez-Sansigre}, A., {Gonz{\'a}lez Delgado}, R., \&
  {P{\'e}rez-Torres}, M. 2010, \mnras, L95+

\bibitem[{{Wampler} {et~al.}(1975){Wampler}, {Burbidge}, {Baldwin}, \&
  {Robinson}}]{wamp75}
{Wampler}, E.~J., {Burbidge}, E.~M., {Baldwin}, J.~A., \& {Robinson}, L.~B.
  1975, \apjl, 198, L49

\bibitem[{{Weilbacher} {et~al.}(2003){Weilbacher}, {Duc}, \&
  {Fritze-v.~Alvensleben}}]{weilbacher03}
{Weilbacher}, P.~M., {Duc}, P., \& {Fritze-v.~Alvensleben}, U. 2003, \aap, 397,
  545

\bibitem[{{Weilbacher} {et~al.}(2002){Weilbacher}, {Fritze-v.~Alvensleben},
  {Duc}, \& {Fricke}}]{weilbacher02}
{Weilbacher}, P.~M., {Fritze-v.~Alvensleben}, U., {Duc}, P., \& {Fricke}, K.~J.
  2002, \apjl, 579, L79

\bibitem[{{Weniger} {et~al.}(2009){Weniger}, {Theis}, \& {Harfst}}]{weniger}
{Weniger}, J., {Theis}, C., \& {Harfst}, S. 2009, Astronomische Nachrichten,
  330, 1019

\bibitem[{{Wild} {et~al.}(2009){Wild}, {Walcher}, {Johansson}, {Tresse},
  {Charlot}, {Pollo}, {Le F{\`e}vre}, \& {de Ravel}}]{wild09}
{Wild}, V., {Walcher}, C.~J., {Johansson}, P.~H., {Tresse}, L., {Charlot}, S.,
  {Pollo}, A., {Le F{\`e}vre}, O., \& {de Ravel}, L. 2009, \mnras, 395, 144

\bibitem[{{Wilhite} {et~al.}(2005){Wilhite}, {Vanden Berk}, {Kron},
  {Schneider}, {Pereyra}, {Brunner}, {Richards}, \& {Brinkmann}}]{wilhite05}
{Wilhite}, B.~C., {Vanden Berk}, D.~E., {Kron}, R.~G., {Schneider}, D.~P.,
  {Pereyra}, N., {Brunner}, R.~J., {Richards}, G.~T., \& {Brinkmann}, J.~V.
  2005, \apj, 633, 638

\bibitem[{{Willmer} {et~al.}(2006){Willmer}, {Faber}, {Koo}, {Weiner},
  {Newman}, {Coil}, {Connolly}, {Conroy}, {Cooper}, {Davis}, {Finkbeiner},
  {Gerke}, {Guhathakurta}, {Harker}, {Kaiser}, {Kassin}, {Konidaris}, {Lin},
  {Luppino}, {Madgwick}, {Noeske}, {Phillips}, \& {Yan}}]{willmer2006}
{Willmer}, C.~N.~A., {et~al.} 2006, \apj, 647, 853

\bibitem[Woods 
\& Geller(2007)]{woods} Woods, D.~F., \& Geller, M.~J.\ 2007, \aj, 134, 527 

\bibitem[{{Yan} {et~al.}(2006){Yan}, {Newman}, {Faber}, {Konidaris}, {Koo}, \&
  {Davis}}]{yan2006}
{Yan}, R., {Newman}, J.~A., {Faber}, S.~M., {Konidaris}, N., {Koo}, D., \&
  {Davis}, M. 2006, \apj, 648, 281

\end{thebibliography}
\end{document}